\documentclass[12pt,a4paper,final]{iopart}
\usepackage{cite}
\usepackage{bm}
\usepackage{float}
\usepackage{stmaryrd,scalerel} 
\usepackage{iopams}
\usepackage{color}
\usepackage[breaklinks=true,colorlinks=true,linkcolor=blue,urlcolor=blue,citecolor=blue]{hyperref}
\expandafter\let\csname equation*\endcsname=\relax
\expandafter\let\csname endequation*\endcsname=\relax
\usepackage{amsmath}

\expandafter\let\csname equation*\endcsname\relax

\expandafter\let\csname endequation*\endcsname\relax

\usepackage{amsmath}

\begin{document}

\title{Ornstein--Uhlenbeck Process Driven by Multiple Dichotomous Noises}
\author{Silvio Kalaj$^{1,2}$, Dongho Lee$^3$,  Enzo Marinari$^{2,4}$, Jae-Hyung Jeon$^{3,5}$, Pascal Viot$^1$ \& Gleb Oshanin$^{1,5}$}
\address{$^1$ Sorbonne Universit\'e, CNRS, Laboratoire de Physique Th\'eorique de la Mati\`ere Condens\'ee (UMR CNRS 7600), 4 Place Jussieu, 75252 Paris Cedex 05, France \\
	$^2$ Dipartimento di Fisica, Sapienza Universit\`a di Roma, P.le A. Moro 2, I-00185, Roma, Italy\\
	$^3$  Department of Physics, Pohang University of Science and Technology (POSTECH), Pohang 37673, Republic of Korea\\
	$^4$ Nanotech-CNR, UOS di Roma and INFN, Sezione di Roma, P.le A. Moro 2, I-00185, Roma,
	Italy\\
$^5$  Asia Pacific Center for Theoretical Physics (APCTP),
Pohang 37673, Republic of Korea}

\date{today}

\begin{abstract}
We study a generalized Ornstein--Uhlenbeck process driven by a superposition of $K$ independent dichotomous noises with arbitrary fixed amplitudes and switching rates. Unlike the classical Ornstein--Uhlenbeck process driven by equilibrium Gaussian white noise, the present system is governed by bounded nonequilibrium fluctuations with finite correlation times. We obtain exact expressions for the stationary position distribution and all cumulants, and show that the stationary state possesses an unexpectedly rich structure, including compact support, algebraic branch-point singularities, edge divergences, and multiple extrema. We establish a mapping onto a heterogeneous random-flight process with bounded jumps, yielding a transparent probabilistic interpretation of the stationary measure. We further analyze several limiting regimes, including the crossover to Gaussian statistics for large numbers of noise sources. For ensembles with exponentially-distributed quenched amplitudes, we derive exact disorder-averaged stationary distributions and show that disorder fundamentally alters the stationary state, producing exponential tails decorated by algebraic prefactors with non-trivial exponents.
	\end{abstract}
		
	
	\section{Introduction}
	
The Ornstein--Uhlenbeck process is one of the fundamental models of stochastic dynamics and has found widespread applications in statistical physics, quantitative biology, and financial mathematics \cite{1,2,3,4,5,BouchaudPotters}. In its classical form, it describes a linear relaxation dynamics driven by Gaussian white noise and constitutes one of the few nontrivial stochastic differential equations that admit a fully explicit solution. In physics, it arises naturally in a variety of contexts, including the motion of overdamped particles confined in optical traps (see, e.g., \cite{6, njp_jeon}) and particles tethered to polymer backbones (see, e.g., \cite{7}). Numerous additional applications have been reviewed in recent works \cite{11,12}.
More generally, within the framework of statistical physics and stochastic thermodynamics \cite{8,9}, the white-noise forcing represents the action of an equilibrium thermal bath, while the linear restoring force accounts for dissipation. Together, these two ingredients establish the fluctuation--dissipation balance characteristic of systems at thermal equilibrium. Owing to its stationary Gaussian distribution, exactly solvable correlation structure, and Markovian nature, the Ornstein--Uhlenbeck process has become a paradigmatic model for fluctuations around equilibrium states.

In many realistic systems, however, the assumption that the driving fluctuations can be represented by Gaussian white noise is merely an idealization of limited validity. While appropriate for systems in thermal equilibrium, it often fails to capture the physical mechanisms responsible for fluctuations in complex environments. In particular, the noise acting on a system is frequently generated by an environment that is itself out of equilibrium and therefore cannot be viewed as an equilibrium thermal bath. 
Biological media~\cite{Libchaber2000, maggi2014, zhao2024}, intracellular environments~\cite{guo2014, jeon2018}, ecological systems~\cite{wienand2017}, and active matter~\cite{caspi2000, tailleur2008, gov2020  }, for instance, typically produce stochastic forces with finite correlation times, persistent temporal correlations, and a finite number of internal states, rather than continuous Brownian agitation. In such situations, the driving fluctuations are more naturally described by finite-state Markov processes or stochastic switching dynamics~\cite{dhar2019, jaegon2019}.
Similar considerations arise in financial mathematics, where volatility, liquidity, and macroeconomic conditions often evolve through abrupt transitions between metastable market regimes rather than by continuous Gaussian fluctuations~\cite{Hamilton89,Elliott05}. More generally, modern stochastic-volatility models increasingly interpret financial noise as the manifestation of complex nonequilibrium environments characterized by multiple intrinsic time scales and long-lived temporal correlations \cite{BouchaudPotters,Bouchaud2003,Perello2004}. Regime-switching diffusions and telegraph-type stochastic processes have consequently emerged as analytically tractable and physically motivated alternatives to purely Gaussian models \cite{Ratanov08,Ratanov07}.

In this work, we study an exactly solvable generalization of the Ornstein--Uhlenbeck process driven by multiple dichotomous noises acting in parallel. Specifically, the stochastic forcing is taken to be $\Omega(t) = \sum_k \omega_k(t)$, where the $\omega_k(t)$ are statistically independent dichotomous noises with arbitrary amplitudes (velocities)  $v_k$ and switching rates $\lambda_k$. In contrast to Gaussian white noise, which models an equilibrium thermal bath, dichotomous noise is an intrinsically nonequilibrium process characterized by a finite correlation time and bounded fluctuations. If the components $\omega_k(t)$ were Gaussian white noises, the resulting dynamics would trivially reduce to the classical Ornstein--Uhlenbeck process with its stationary Gaussian distribution. The behavior under dichotomous driving is fundamentally different. Although the dynamics remains linear, the superposition of independent dichotomous noises produces a bounded nonequilibrium forcing with strongly non-Gaussian statistics.
As a consequence, the stationary position probability density function (PDF) exhibits a remarkably rich structure. Its support is confined to a finite interval and, depending on the amplitudes and switching rates of the constituent noises, it may display multiple local maxima and minima, algebraic singularities, and discontinuities of its derivatives. 
These features are absent in the conventional Ornstein--Uhlenbeck process and demonstrate that even linear relaxation dynamics can generate highly nontrivial stationary states when driven by finite-state nonequilibrium noises.

The stochastic process studied here has several interpretations across different disciplines. At a general level, it describes a linear relaxation dynamics driven by a collection of independent nonequilibrium switching processes acting on distinct time scales. Such a structure emerges whenever a system is subject to multiple fluctuating inputs that alternate between discrete states and contribute additively to the overall forcing. Therefore, 
a natural application concerns the dynamics of an optically trapped tracer particle embedded in a heterogeneous medium and subjected to a superposition of telegraphic fluctuations with different amplitudes and switching rates. Situations of this type may arise in crowded biological or soft-matter systems, where surrounding constituents generate fluctuations spanning a broad range of temporal and spatial scales. Within this picture, rapidly switching low-amplitude components may be associated with small mobile constituents, whereas slowly switching high-amplitude components effectively represent the influence of larger structures, such as vesicles, organelles, or collective rearrangements of the surrounding medium. The resulting dynamics thus provides a simple coarse-grained description of tracer motion in a hierarchically organized or multicomponent random environment.
We note parenthetically that relaxation dynamics of Ornstein--Uhlenbeck processes in heterogeneous media have also been considered \cite{OUFD1, OUFD2}. 
In these models, however, environmental heterogeneity is represented by temporal fluctuations of the diffusivity, whereas here it enters through additive nonequilibrium forcing.

From the perspective of financial mathematics, the present model provides a natural framework for constructing a new class of exactly solvable stochastic-volatility models. Since the square of the Ornstein--Uhlenbeck process is non-negative by construction, it can be naturally interpreted as a stochastic variance (or diffusivity) process. In this way, our approach extends the model introduced in \cite{lee}, where the variance was taken to be the square of an Ornstein--Uhlenbeck process driven by a single dichotomous noise. Here, the driving consists instead of a superposition of independent dichotomous noises with arbitrary amplitudes and switching rates. Consequently, the variance process exhibits several intrinsic time scales associated with the different switching mechanisms, while remaining fully analytically tractable.
This construction may be viewed as an exactly solvable nonequilibrium counterpart of the Cox--Ingersoll--Ross (CIR) model \cite{Cox1985}. Unlike the classical CIR model, where the variance is driven by a single Gaussian source, the present model is governed by multiple independent finite-state stochastic drivers. It therefore naturally captures features commonly encountered in financial time series - including non-Gaussian fluctuations, intermittency, and a hierarchy of temporal correlation times - while retaining the analytical transparency of an exactly solvable model.

The paper is organized as follows. In Sec.~\ref{model}, we introduce the model and define its principal parameters. In Sec.~\ref{integral}, building on the exact solution of the Ornstein--Uhlenbeck process driven by a single dichotomous noise \cite{sancho,we}, we derive the stationary position PDF for an arbitrary finite collection of independent dichotomous noises. We present both integral and Fourier-series representations of the solution and introduce an equivalent random-flight model that provides a simple probabilistic interpretation of the process. In Sec.~\ref{asymp}, we investigate the bulk properties of the stationary PDF by deriving its complete hierarchy of cumulants and constructing the corresponding Edgeworth expansion. We also analyze the asymptotic behavior of the PDF near the boundaries of its compact support. In Sec.~\ref{random}, we consider the case of quenched disorder, in which the noise amplitudes are independent exponentially distributed random variables, and determine the corresponding disorder-averaged stationary PDF. Finally, in Sec.~\ref{concs}, we summarize our main results and discuss possible extensions and applications of the present framework.
Lengthy technical calculations are relegated to the Appendices. In~\ref{C}, we derive the cumulants of the stationary position distribution to arbitrary order, while~\ref{B} is devoted to the asymptotic analysis of the large-$|x|$ tails of the disorder-averaged position $x$ PDF for exponentially distributed quenched noise amplitudes.

\section{Model}
\label{model}

Consider a one-dimensional stochastic differential equation of the  form
\begin{align}
	\label{LOU}
	\dot{x}(t) = - \gamma x(t) + \Omega(t) \,, \quad x(0) = 0 \,, \quad 0 \leq t  < \infty \,,
	\end{align}
	where the dot denotes the time derivative, $x(t)$ is an instantaneous position, $\gamma > 0$ is the relaxation rate, while 
	\begin{align}
		\Omega(t)  = \sum_{k=1}^K \omega_k(t) \,,
		\end{align}
		with $\omega_k(t)$ being the independent dichotomous noises that switch randomly between  the  values $\pm v_k$ with (in general, unequal) switching rates $\lambda_k$. In eq. \eqref{LOU}, the stochastic variable $x(t)$ has units of length, the relaxation rate $\gamma$ has units of $1/{\rm time}$, while the driving process $\Omega(t)$ and the amplitudes $v_k$ have units of velocity.
		
		The covariance function of each noise obeys  \cite{hanggi}
		\begin{align}
E_k\left\{\omega_k(t) \omega_k(t')\right\} = v_k^2 \exp\left(- 2 \lambda_k  |t - t'|\right) = \frac{D_k}{\tau_k} \exp\left(- \frac{|t - t'|}{\tau_k}\right) \,,
			\end{align}
			where the symbol 
			$E_k\left\{\ldots\right\}$  here and henceforth denotes averaging with respect to different realizations of the $k$-th dichotomous  noise, 	
			\begin{align}
	\label{diff}
				\tau_k = \frac{1}{2 \lambda_k} \quad \text{and} \quad D_k = \frac{v_k^2}{2 \lambda_k} \,.
				\end{align}
Recall that the process is non-Gaussian and therefore  
does not satisfy Wick's theorem. Although all odd-order 
correlation functions identically vanish by symmetry, even-order correlations do not factorize into products of two-point functions. 
Consequently, higher-order cumulants are nonzero, reflecting the intrinsically non-Gaussian character of dichotomous noise. 
It is also worth noting that the dichotomous noise converges to Gaussian white noise in the limit $\tau_k \to 0$ with $D_k$ (which has the meaning of the diffusion coefficient) in eq. \eqref{diff} held constant \cite{hanggi}, i.e., by taking $v_k \to \infty$ and $\lambda_k \to \infty$ simultaneously, while keeping their ratio $D_k$ in eq. \eqref{diff}  fixed. 

\begin{figure}[t]
	\begin{center}
		\includegraphics[width=155mm]{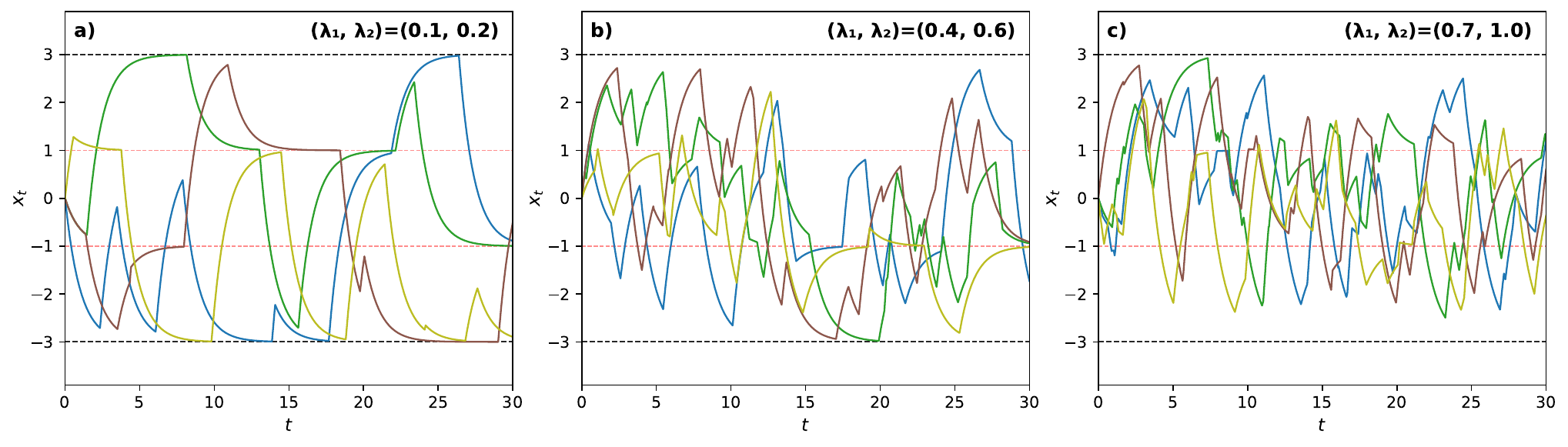}
	\end{center}
	\caption{Sample trajectories of the process $x(t)$ defined in eq.~\eqref{LOU} with $K=2$, $\gamma=1$, and velocities $v_1=1$ and $v_2=2$. 
    From left to right, the switching rates are $(\lambda_1,\lambda_2)=(0.1,0.2)$, $(0.4,0.6)$, and $(0.7,1.0)$.
    The horizontal red dashed lines indicate the inner branch points $x=\Omega/\gamma=\pm1$, while the black dashed lines depict the boundaries of the support of the process, $x=\pm\Omega_{\rm max}/\gamma=\pm3$.
}
	\label{fig:1} 
	\end{figure}

In Fig.~\ref{fig:1}, we present sample trajectories of the process defined by eq.~\eqref{LOU} for $K=2$ and $\gamma=1$. 
The velocities are fixed at $(v_1,v_2)=(1,2)$, while the switching rates increase from left to right as $(\lambda_1,\lambda_2)=(0.1,0.2)$, $(0.4,0.6)$, and $(0.7,1.0)$. 
For the smallest switching rates [panel (a)], the trajectories exhibit pronounced plateaus near the values indicated by the red and black dashed lines. 
This occurs because the time intervals between successive switches are sufficiently long for the process to relax toward the branch point $x=\Omega/\gamma$ associated with a given state of the stochastic forcing. 
In the present case, the forcing $\Omega(t)=\omega_1(t)+\omega_2(t)$ can take the four possible values $\Omega=\pm v_1\pm v_2=\pm1,\pm3$. 
The maximum forcing amplitude, $\Omega_{\rm max}=v_1+v_2=3$, sets the boundaries of the support of $x(t)$, i.e., $|x(t)|\leq \Omega_{\rm max}/\gamma$.
As the switching rates increase, the plateaus become less pronounced and the trajectories exhibit more rapid fluctuations. 
Nevertheless, the boundaries remain unchanged as they are determined solely by the velocities for fixed $\gamma$.

			\section{Position probability density function  in the stationary state}
			\label{integral}
			
We focus here on the general expression for the position PDF $P(x)$ in the stationary state, i.e., for $t = \infty$. To this end, we determine first the stationary  state characteristic 
function, 
\begin{align}
	\label{m}
	\Phi(\nu)= \int^{\infty}_{-\infty} dx \, e^{i \nu x} \, P(x)  \,.
	\end{align}
Alternatively, $\Phi(\nu)$ can be written as 
\begin{align}
	\Phi(\nu) = \lim_{t \to \infty}\prod_{k=1}^K E_k\left\{e^{i \, \nu \, x_k(t)}\right\} \,,
	\end{align}
where $x_k(t)$ is the solution of eq. \eqref{LOU} for a given realization of noises, 
\begin{align}
	x_k(t) = \int^t_0 d\tau \, e^{- \gamma  (t- \tau)} \, \omega_{k}(\tau)   \,.
	\end{align}
Consequently, the stationary state characteristic function attains the form
\begin{align}
	\label{Phi}
	\Phi(\nu) = \prod_{k=1}^K {\cal Z}_k(\beta_k \nu) =  \lim_{t \to \infty} \prod_{k=1}^K E_k\left\{\exp\left(i \, \nu \, \int^t_0 d\tau \, e^{- \gamma  (t- \tau)} \, \omega_k(\tau) \right)\right\} \,.
	\end{align}
From \cite{we} we have
\begin{align}
	\begin{split}
		\label{Z}
	{\cal Z}_k(\beta_k \nu) &= \lim_{t \to \infty} E_k\left\{\exp\left(i \, \nu \, \int^t_0 d\tau \, e^{- \gamma  (t- \tau)} \, \omega_k(\tau) \right)\right\} \\   
	&= 2^{\alpha_k - 1/2} \Gamma(\alpha_k + 1/2) \left(\beta_k \nu\right)^{1/2 - \alpha_k} J_{\alpha_k - 1/2}(\beta_k \nu) \,,
	\end{split}
	\end{align}
where $J_r(z)$ is the Bessel function of the first kind  \cite{WATSON}, while $\alpha_k$ and $\beta_k$ are the reduced switching rates and reduced (velocities) amplitudes, respectively, 
\begin{align}
	\alpha_k = \lambda_k/\gamma > 0 \,, \quad \beta_k = v_k/\gamma > 0 \,.
	\end{align}
	Therefore, the stationary state characteristic function is completely defined.
	
\subsection{Integral representation of the position PDF}

In virtue of eqs. \eqref{m} and \eqref{Z}, the integral representation of the desired position PDF is given  by the following Fourier integral
\begin{align}
	\label{PDF}
	P(x) = \frac{1}{\pi} \int^{\infty}_0 d\nu \, \cos(\nu x) \, \prod_{k=1}^K {\cal Z}_k(\beta_k \nu) \,.
	\end{align} 
The above position PDF is manifestly normalized to unity, since each ${\cal Z}_k(\beta_k \nu)=1$ for $\nu=0$.
Furthermore, the PDF possesses several remarkable analytic properties. First, likewise for $K = 1$  \cite{sancho} and for $K = 2$ \cite{we} cases, $P(x)$ has compact support and vanishes
\begin{align}
	\label{BB}
P(x)\equiv 0 \qquad \text{for} \, |x|>B, \quad B=\sum_{k=1}^K \beta_k.
\end{align}
This is, of course, not counter-intuitive because the process $x(t)$ evolves on a bounded interval, as discussed above. 
Second, within the interval $[-B,B]$, the PDF is not analytic everywhere but possesses a finite set of algebraic branch-points $x_p$ \cite{WATSON,URSELL}, defined as signed sums of the parameters $\beta_k$; that being, 
\begin{align}
	\label{branch}
	x_p=\sum_{k=1}^K \epsilon_k \beta_k,
\qquad \epsilon_k = \pm 1 \,.
\end{align} 
These $2^K$ branch-points $x_p$, $p= 1,2, \ldots, 2^K$ (counted with multiplicity, since some may coalesce when the $\beta_k$ are rationally related) correspond to different choice of variables $\epsilon_k$ and define
the values of $x$ at which the multidimensional stationary-phase contour in the Sommerfeld representation of the Bessel product becomes pinched, resulting in a multivalued analytic continuation of $P(x)$.
Near each such point $x_p$, the PDF exhibits the local behavior of the form
\begin{align}
P(x)\simeq |x-x_p|^{\mu_p} \,,
\end{align}
with some exponent $\mu_p>-1$, ensuring integrability. Here and throughout the paper, the symbol ``$\simeq$'' denotes the leading asymptotic behavior, with multiplicative prefactors and higher-order correction terms left unspecified.
The value of $\mu_p$ depends on the multiplicity of coinciding sums $\sum_k \epsilon_k \beta_k$ and on the orders $\alpha_k-1/2$ of the Bessel functions. Depending on the value of $\mu_p$, these branch-points correspond to locations where the PDF changes convexity or attains extremal values, indicating the points where trajectories either accumulate or, conversely, are comparatively unlikely to visit.
Note that the endpoints of the support, $x=\pm B$, are special cases of the branch-points, where the PDF may remain finite, become constant, or even diverge. Behavior at the branch-points for the case $K=2$ has recently been discussed in a different context in ~\cite{we}. Below we will consider the behavior at the endpoints of the support for arbitrary $K$.

We close this subsection by analyzing the position PDF in the diffusion limit introduced below eq. \eqref{diff}. Assuming that the amplitudes and the switching rates of all $K$ components of noise $\Omega(t)$ tend to infinity, while $D_k$ are kept fixed, we have 
\begin{align}
	\lim_{v_k, \lambda_k \to \infty} \ln {\cal Z}_k(\beta_k \nu) = - \frac{D_k}{2 \gamma} \nu^2 \,, 
	\end{align}
and consequently, 	
	\begin{align}
	\prod_{k=1}^K {\cal Z}_k(\beta_k \nu) = \prod_{k=1}^K \exp\left(\ln {\cal Z}_k(\beta_k \nu)\right) = \exp\left( - \left(\sum_{k=1}^K D_k\right) \frac{\nu^2}{2 \gamma}\right) \,.
	\end{align}
Therefore, in the diffusion limit, the combined dichotomous driving becomes equivalent to an effective Gaussian white noise, and the dynamics reduces to that of a standard Ornstein–Uhlenbeck process. As a consequence, the stationary position PDF assumes the familiar Gaussian form, with an effective diffusion coefficient given by the sum of the diffusion coefficients associated with the individual dichotomous noise components.
In Sec. \ref{asymp}, we will return to this issue and study, in a general setting with arbitrary amplitudes and switching rates, how the exact stationary PDF in eq. \eqref{PDF} approaches its Gaussian limit. To this end, we will construct the corresponding Edgeworth expansion, which provides a systematic characterization of deviations from Gaussian behavior, and also 
derive exact expressions for all cumulants of the position PDF.

\subsection{An equivalent random flight model}
\label{RF}

\begin{figure}[h]
	\begin{center}
		\includegraphics[width=155mm]{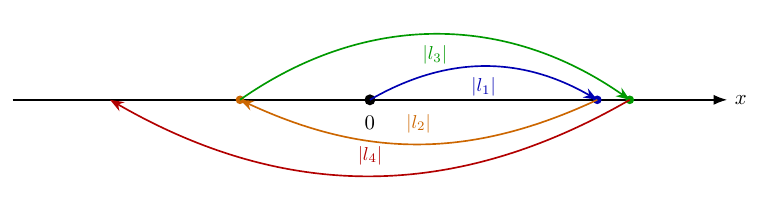}
	\end{center}
	\caption{An equivalent random-flight model. The displacements $l_k$ are independently sampled quenched random variables, each drawn from the distribution given in eq.~\eqref{plk}; the realization chosen at a given step is independent of those selected at all previous and subsequent steps.}
	\label{fig:2} 
\end{figure}

 The PDF defined in eq.~\eqref{PDF} also admits a simple probabilistic interpretation. Consider a $K$-step random walk (see Fig.~\ref{fig:2} for a schematic illustration) in which, at the $k$-th step, a random displacement $l_k$
 is drawn independently of all previous  or subsequent steps from a symmetric probability density $p(l_k)$ supported on the interval $[-\beta_k,\beta_k]$:
\begin{align}
	\begin{split}
	\label{plk}
		p(l_k) &=  C_k \begin{cases}
		\left(\beta_k^2-l_k^2\right)^{\alpha_k-1}\,,   
		\qquad |l_k|\leq \beta_k,\\
		0 \,, \qquad \qquad \qquad \quad  |l_k| > \beta_k\,,\\ 
			\end{cases}\\
			C_k &= \frac{\Gamma(\alpha_k+1/2)}
			{\sqrt{\pi} \,\Gamma(\alpha_k)\,\beta_k^{2\alpha_k-1}}
\end{split}
\end{align}
Positive values of $l_k$ correspond to flights to the right, while negative values correspond to flights to the left. The final position after $K$ steps is then $x = \sum_{k=1}^K l_k$.

Such a  random walk is 
strongly heterogeneous with the statistics of the 
jump lengths dependent on the parameter 
$\alpha_k$: For $\alpha_k > 1$, short jumps are 
favored, whereas for $\alpha_k = 1$ the jump length is 
uniformly distributed over the interval 
$[-\beta_k,\beta_k]$. 
In contrast, when $\alpha_k < 1$, the distribution becomes peaked near the extremal values $\pm \beta_k$, thereby favoring long jumps.
Following the standard analysis of random walks based on characteristic functions (see, e.g., \cite{hugues}), one finds that the position PDF of such a walk after $K$ steps is given precisely by eq.~\eqref{PDF}. 
	
\subsection{Series representation of the position probability density function}
\label{series}

The PDF defined by the oscillatory integral in eq.~\eqref{PDF}, which involves a product of Bessel functions, is generally difficult to evaluate numerically; namely, the integrand becomes strongly oscillatory for large values of the Fourier variable $\nu$ and the product of several Bessel functions may produce substantial numerical cancellations. As a consequence, direct numerical integration becomes increasingly cumbersome and potentially unstable when the number $K$ of dichotomous components increases or when the parameters $\alpha_k$  and $\beta_k$ span widely different scales. It is therefore natural to seek alternative representations that are more suitable for numerical computations.

We follow the approach developed in 
\cite{bennett,barakat} for the Rayleigh-flight model
   in two dimensions (see, e.g., \cite{hugues}), in which the position PDF has been obtained in form of the Fourier-Bessel series. Here 
    we also exploit the fact that 
   $P(x)$ has compact support and derive equivalent
   representation of the PDF in eq.~\eqref{PDF} in
   the form of Fourier series on the interval $[-B,B]$ (see eq. \eqref{BB}). Such a
       representation can be truncated at arbitrary order, 
       allowing one to reconstruct reliably well the PDF without
        performing delicate integrations of oscillatory functions.

We formally expand $P(x)$ in eq. \eqref{PDF}, which is an even function of $x$, in the Fourier cosine series on the interval $x \in [-B,B]$,
\begin{align}
	\label{f}
	P(x) = \frac{c_0}{2} +\sum_{n=1}^{\infty} c_n \cos\left(\frac{\pi n x}{B}\right) \,,\quad c_n = \frac{2}{B} \int^B_0  dx \, P(x) \, \cos\left(\frac{\pi n x}{B}\right) \,,
	\end{align}
	and seek to determine the coefficients in this expansion. 

Since $P(x)$ vanishes identically for $|x| > B$, the upper limit of integration may be extended to infinity without changing the value of the integral. Hence, 
\begin{align}
	\begin{split}
		\label{def}
		c_n
		&=\frac{2}{B}\int_0^\infty dx \, P(x)
		\cos\left(\frac{\pi n x}{B}\right) \\
		&=\frac{2}{\pi B}\int_0^\infty dx \,
		\cos\left(\frac{\pi n x}{B}\right)
		\int_0^\infty d\nu \, \cos(\nu x)
		\prod_{k=1}^K{\cal Z}_k(\beta_k\nu),
	\end{split}
\end{align}
where the second line follows directly from the Fourier-cosine representation of $P(x)$, eq.~\eqref{PDF}.
To proceed, we invoke the Fourier cosine inversion theorem. Specifically, if $f(x)$ is absolutely integrable on $[0,\infty)$ and satisfies the standard regularity conditions ensuring Fourier inversion, then
\begin{align}
	\label{inv}
	f(x) = 
		\frac{2}{\pi}
	\int_0^\infty d\nu \,\cos(\nu x)
	\int_0^\infty dy \, \cos(\nu y) \, f(y).
\end{align}
Comparing the above expression and the expression in the second line in eq. \eqref{def},  we infer that
\begin{align}
	\label{cn}
	c_n = \frac{1}{B} \prod_{k=1}^K {\cal Z}_k\left(\beta_k \frac{\pi n}{B}\right) \,,
	\end{align}
which provides the desired definition of the coefficients in the Fourier series \eqref{f}.

\begin{figure}[htbp]
	\begin{center}
		\includegraphics[width=155mm]{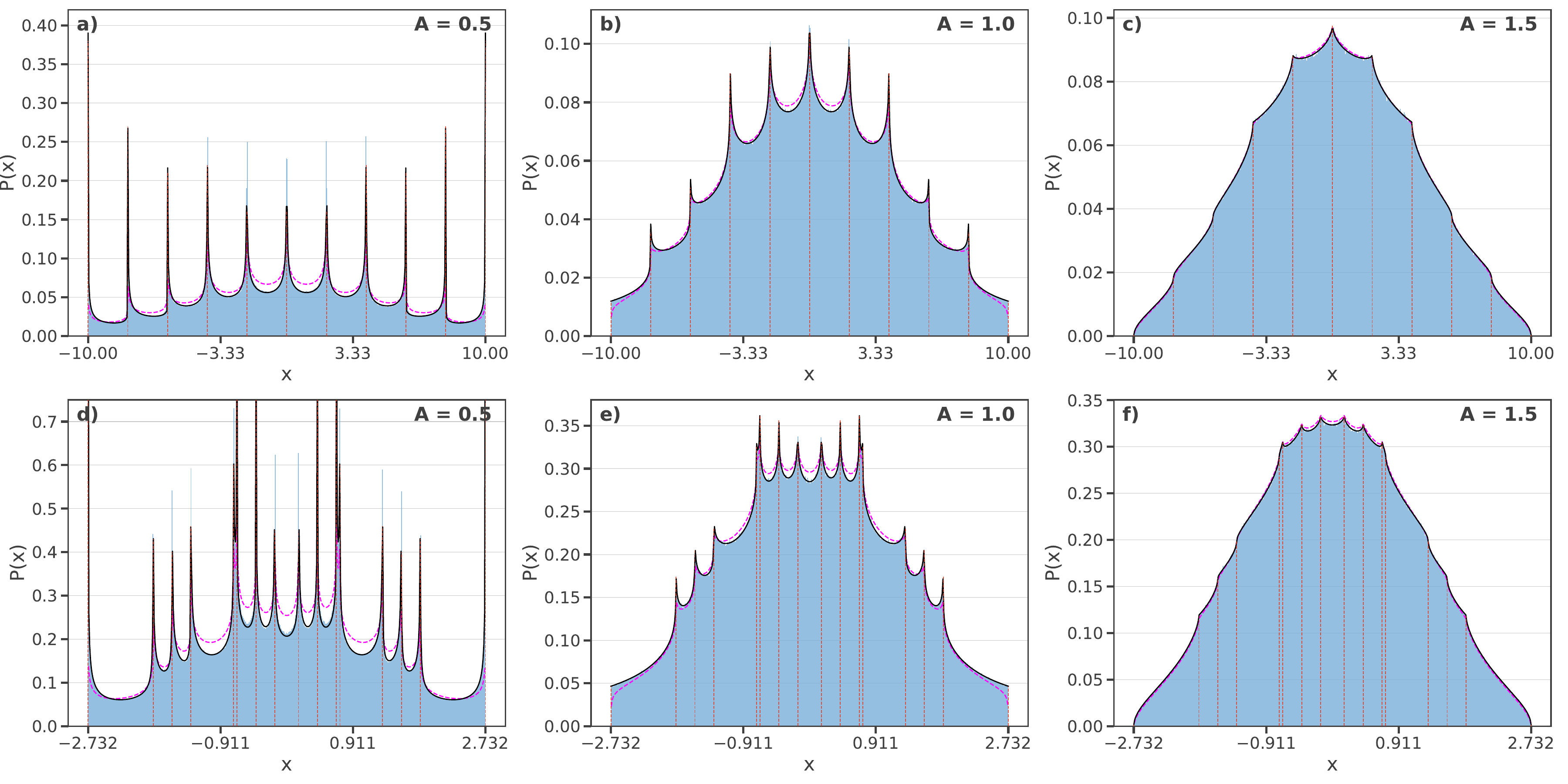}
	\end{center}
	\caption{Position probability density function $P(x)$ of the process defined by eq.~\eqref{LOU}, driven by a superposition of four independent dichotomous noises. The first row corresponds to reduced velocities $\beta_k=k$, $k= 1, 2, 3, 4$, while in the second row $\beta_1=1$, $\beta_2=1/\sqrt{2}$, $\beta_3=1/\sqrt{3}$, and $\beta_4=1/\sqrt{5}$.
		The vertical red dashed lines indicate the locations of the branch points. The dashed magenta curves are obtained by numerical evaluation of the integral representation in eq.~\eqref{PDF}, while the solid black curves correspond to the Fourier series representation given by eqs.~\eqref{f} and \eqref{cn}, truncated after $10^3$ terms.
			The panels differ by the values of the reduced switching rates: $\alpha_1=\alpha_2=0.1$ and $\alpha_3=\alpha_4=0.15$ (panels (a), (d); $A<1$, eq.~\eqref{A}); $\alpha_1=\alpha_2=0.2$ and $\alpha_3=\alpha_4=0.3$ (panels (b), (e); $A=1$); and $\alpha_1=\alpha_2=0.3$, $\alpha_3=0.4$, and $\alpha_4=0.5$ (panels (c), (f); $A>1$).
		Histograms: empirical distributions obtained from $10^7$ realizations of the random-flight process introduced in Subsec.~\ref{RF}.}
	\label{fig:3} 
\end{figure}

Figure~\ref{fig:3} presents the stationary position PDF $P(x)$ of the process defined by eq.~\eqref{LOU}, driven by a superposition of four independent dichotomous noises. The figure compares results obtained from numerical simulations of the equivalent random-flight model (averaged over $10^7$ realizations) with our analytical predictions. Specifically, the dashed magenta curves correspond to the Fourier integral representation in eq.~\eqref{PDF} evaluated numerically using Python, while the solid black curves are obtained from the Fourier series representation in eqs.~\eqref{f} and \eqref{cn}, truncated after $10^3$ terms. While both analytical representations describe the same probability density function, their numerical performance differs significantly. The Fourier-series representation yields results that are essentially indistinguishable from the simulations, whereas the direct numerical evaluation  of the oscillatory Fourier integral using Python exhibits noticeable inaccuracies in between of the singular points. Nevertheless, the latter still reproduces the overall shape and main features of the distribution.

The first row in Fig.~\ref{fig:3}  corresponds to reduced velocities $\beta_k=k$, $k = 1, 2, 3, 4$, while the second row corresponds to the choice $\beta_1=1$, $\beta_2=1/\sqrt{2}$, $\beta_3=1/\sqrt{3}$, and $\beta_4=1/\sqrt{5}$, respectively. The vertical red dashed lines indicate the corresponding locations of the branch-points (see eq. \eqref{branch}). 
The different panels in Fig.~\ref{fig:3} correspond to different values of the reduced switching rates: $\alpha_1=\alpha_2=0.1$ and $\alpha_3=\alpha_4=0.15$ (panels (a) and (d)); $\alpha_1=\alpha_2=0.2$ and $\alpha_3=\alpha_4=0.3$ (panels (b) and (e)); and $\alpha_1=\alpha_2=0.3$, $\alpha_3=0.4$, and $\alpha_4=0.5$ (panels (c) and (f)). Histograms represent the results of $10^7$ realizations of the random-flight process introduced in Subsec.~\ref{RF}.

Let us define the sum of $\alpha_k$ as
\begin{align}
	\label{A}
	A = \sum_{k=1}^K \alpha_k \,.
\end{align}
Several qualitatively distinct behaviors emerge depending on the value of the parameter $A$, as can already be discerned from Fig.~\ref{fig:1}, where we deliberately show several sample trajectories for $A<1$ (panel (a)), $A=1$ (panel (b)), and $A>1$ (panel (c)). For $A<1$ (panels (a) and (d), Fig.~\ref{fig:3}), the PDF is composed of a sequence of $U$-shaped segments located between neighboring branch points. The density diverges algebraically at each branch point and, in addition, exhibits algebraic divergences at the edges of its support (see Subsec.~\ref{edges}). Consequently, all peaks visible in panels (a) and (d) in Fig.~\ref{fig:3} are in fact singular and attain infinite height. In this regime, the PDF possesses a highly nontrivial structure, with local maxima located at the branch points and local minima between them.

The intermediate case $A=1$, shown in panels (b) and (e) in Fig.~\ref{fig:3}, retains much of the structure observed for $A<1$. The PDF is still composed of successive $U$-shaped pieces separated by branch points. However, the singularities are now marginal, and the peaks acquire finite heights (see Subsec.~\ref{edges} for a detailed discussion). In panel (b) the global maximum occurs at the origin, while the heights of other maxima decrease progressively away from the center. In contrast, panel (e) exhibits several maxima of identical height in the vicinity of the origin. Another notable difference concerns the behavior at the boundaries of the support: whereas the PDF diverges at the edges for $A<1$, it approaches finite nonzero values when $A=1$.

A qualitatively different situation arises for $A>1$ (panels (c) and (f) in Fig.~\ref{fig:3}). In this regime the PDF becomes a continuous, centrally peaked function that vanishes at the edges of the support. Although the branch-points remain visible, they no longer correspond to divergences of the PDF itself. Instead, they manifest themselves through nonanalytic features, appearing as finite jumps in the derivatives. The resulting distribution is much smoother than in the previous cases, while still retaining a rich multi-peak structure inherited from the underlying branch point singularities.

\section{Behavior of the position PDF in the bulk and at the boundaries of the support}
\label{asymp}

In this Section we analyse the behavior of the PDF both in the bulk of its support and in the vicinity of its boundaries. 

\subsection{Behavior in the bulk of the support}

The position PDFs for $K = 4$ presented in Fig.~\ref{fig:3} differ markedly from a Gaussian distribution, irrespective of whether $A<1$, $A=1$, or $A>1$. This behavior is not particularly surprising in view of the random-flight interpretation discussed above. Indeed, even the classical Rayleigh flight with fixed step lengths is known to exhibit highly non-Gaussian and singular endpoint distributions when the number of steps is not big enough \cite{hugues}. Nevertheless, just as the endpoint distribution of a Rayleigh flight converges to a Gaussian form as the number of steps increases, one may expect that the PDF in eqs.~\eqref{PDF} and \eqref{f} becomes asymptotically Gaussian in the bulk of its support when $K$ becomes sufficiently large. The question, however, is under which conditions such a Gaussian regime emerges and what role is played by the parameters $\alpha_k$ and $\beta_k$. These issues are addressed below.

Note first that for large values of $\nu$, the kernel exhibits the asymptotic power-law behavior
 \begin{align}
 	\label{as}
Z(\nu) = \prod_{k=1}^K {\cal Z}_k(\beta_k \nu) = O\left(\frac{1}{\nu^{A}}\right)\,. 	
 	\end{align}
Therefore, as $A$ increases, the contribution of large values of $\nu$ should become progressively suppressed, implying that the integral in eq.~\eqref{PDF} should be increasingly dominated by the behavior of the integrand in the vicinity of $\nu=0$. In the limit of  small $\nu$, we have
\begin{align}
	\label{kappa2}
\prod_{k=1}^K {\cal Z}_k(\beta_k \nu)\simeq \exp\left(- \frac{ \kappa_2 \nu^2}{2}\right)	\,, \quad \kappa_2 = \sum_{k=1}^K \frac{\beta_k^2}{1 + 2 \alpha_k} \,, 
	\end{align}
	where $\kappa_2$ is the second cumulant (the variance).
	Inserting the above expression into eq. \eqref{PDF} and integrating, one finds that the PDF becomes
	\begin{align} 
		P(x) \simeq \frac{1}{\sqrt{2 \pi \kappa_2}} \exp\left( - \frac{x^2}{2 \kappa_2}\right) \,.
		\end{align}
Therefore, 
the PDF in the bulk of the  support should become a Gaussian function for sufficiently large values of $A$, which is not, of course, a counterintuitive behavior. 
		
We next go one step further and consider the Edgeworth series expansion of the PDF, which permits us to access the rate (and conditions) of convergence to the Gaussian distribution.  Unlike the central limit theorem, which only guarantees weak convergence to a Gaussian limit, the Edgeworth expansion provides an explicit asymptotic correction to the normal density in powers of the inverse sample size (or, more generally, in terms of higher-order cumulants). These corrections generally quantify the effect of skewness, kurtosis, and higher-order cumulants on finite-sample deviations from Gaussianity, although such effects only become fully systematic once one proceeds sufficiently far in the expansion. Here, we restrict attention to the first two terms, which already capture the leading-order non-Gaussian corrections. Using the standard procedure, we get
		\begin{align}
			\label{edgeworth}
			P(x) = \frac{1}{\sqrt{2 \pi \kappa_2}} \exp\left( - \frac{x^2}{2 \kappa_2}\right) \left[1 + \frac{\kappa_4}{24 \kappa_2^2} H_4\left(\frac{x}{\sqrt{\kappa_2}}\right) + \ldots \right] \,,
			\end{align}
			where $H_4(z) = z^4 - 6 z^2 + 3$ is the fourth Hermite polynomial, while $\kappa_4$ is the fourth cumulant of the position probability density, which is given explicitly  by
		\begin{align}
			\label{kappa4}
			\kappa_4 = - 6 \sum_{k=1}^K \frac{\beta_k^4}{\left(1 + 2 \alpha_k\right)^2\left(3 + 2 \alpha_k\right)} \,.
			\end{align} 
The terms omitted from eq. \eqref{edgeworth} involve higher even-order Hermite polynomials of variable $x/
\sqrt{\kappa_2}$, multiplied by combinations of higher even-order cumulants normalized by the corresponding powers of $\kappa_2$. In fact, for the dynamic model studied here all these correction terms can be written  in an explicit form; the cumulants of odd order vanish, while the even ones obey  (see \ref{C} for the details of the derivation), 
\begin{align}
	\label{kappan}
	\kappa_{2 n} = (-1)^{n+1} \frac{(2 n)!}{n} \sum_{k=1}^K \beta_k^{2 n} \sigma_{2 n}(\alpha_k - 1/2) \,,
	\end{align}
where $\sigma_{2 n}(z)$ are the Rayleigh functions (see \eqref{rayleigh} and \cite{kishore} for more details), defined as infinite sums of the reciprocals of even powers of the Bessel function zeros. These functions can also be obtained from the Kishore's recursion relation \cite{kishore}
\begin{align}
	\label{kappan1}
	\sigma_{2n}(z) = \frac{1}{n + z} \sum_{m=1}^{n-1} \sigma_{2 m}(z) \sigma_{2 n- 2m}(z) \,, \quad \sigma_2(z) = \frac{1}{4 (1 + z)} \,,
	\end{align}
We concentrate next on the excess kurtosis $\gamma_4 = \kappa_4/\kappa_2^2$, which obeys for any sequences of $\alpha_k$ and $\beta_k$, the double-sided inequality
\begin{align}
	\label{ineq}
-2 	\leq \gamma_4 < 0 \,. 
	\end{align}   			
The upper bound is immediate, while the lower bound can also be established quite straightforwardly.
The fact that $\gamma_4<0$ implies that the distribution $P(x)$ is always platykurtic, i.e., its tails are systematically lighter than those of a Gaussian distribution with the same variance.
The magnitude  of $|\gamma_4|$ quantifies the fourth-order deviation from Gaussianity, and its decay toward zero provides a measure of the rate of convergence to the Gaussian limit.

To obtain more quantitative insights into the rate of convergence (or not) toward Gaussian function, it is useful to examine several specific situations:
\begin{itemize}
	\item Suppose that all $\alpha_k$ are much smaller than $1$, while $K$ is very large. If all $\beta_k$ are comparable and the limits
	\begin{align}
		\frac{1}{K} \sum_{k=1}^K \beta^2_k = \langle \beta^2\rangle \,, \quad \frac{1}{K} \sum_{k=1}^K \beta^4_k = \langle \beta^4\rangle \,,
		\end{align}
		exist when $K \to \infty$, one has
		\begin{align}\label{Ed_case1}
			\gamma_4 \simeq - 2\frac{ \langle \beta^4\rangle}{\langle \beta^2\rangle^2} \frac{1}{K} \to 0 \,, 
			\end{align}
which signifies that in such a situation the excess kurtosis vanishes with the growth of $K$. If, on the contrary, $\beta_k$ are disproportionally different, e.g., one value of $\beta_k$ completely dominates the sums, the excess kurtosis remains finite and close to the lower bound in eq. \eqref{ineq}.
\item Suppose next that all $\alpha_k$ are much greater than $1$ and the minimal among them is $\alpha_{min}$. The number $K$ of dichotomous noise can be arbitrary. Then, we have
\begin{align}\label{Ed_case2}
	|\gamma_4| \approx 3 \dfrac{\sum_{k=1}^K \beta_k^4/\alpha_k^3}{\left(\sum_{k=1}^K \beta_k^2/\alpha_k\right)^2} \leq 3 \left(\dfrac{\sum_{k=1}^K (\beta_k^2/\alpha_k)^2}{\left(\sum_{k=1}^K \beta_k^2/\alpha_k\right)^2}\right) \frac{1}{\alpha_{min}} \,,
	\end{align} 
i.e., $|\gamma_4|$ is bounded from above by a term including $1/\alpha_{min}$ and therefore, can become arbitrarily small when the reduced switching rate $\alpha_{min}$ increases. 
Also, note that the term in brackets in the right-hand-side of the above expression is always less than $1$.
Moreover, it decreases with $K$ whenever adding new terms continues to contribute a non-negligible fraction of the total sum $\sum_{k=1}^K \beta^2_k/\alpha_k$.  
This and the preceding cases substantiate the discussion below eq. \eqref{as}. 
\item  Suppose now that $K$ is large, while $\alpha_k$ and $\beta_k$ in each realization are drawn from some probability distributions such that the non-zero limits
\begin{align}
	\label{limm}
\lim_{K \to \infty} \frac{1}{K} \sum_{k=1}^K \frac{\beta_k^2}{1+2 \alpha_k} = R_1 ,
\qquad
\lim_{K \to \infty} \frac{1}{K} \sum_{k=1}^K \frac{\beta_k^4}{(3 + 2 \alpha_k) (1+2 \alpha_k)^2} = R_2 ,
\end{align}
exist. Then, using eqs.~\eqref{limm}, one obtains
\begin{align}
\gamma_4 \simeq - \frac{6 R_2}{R_1^2} \frac{1}{K}.
\end{align}
Hence, the excess kurtosis decays universally as $K^{-1}$, independently of the detailed form of the distributions of $\alpha_k$ and $\beta_k$. The only dependence on these distributions enters through the prefactor $6R_2/R_1^2$.
This result admits a simple interpretation. Since both $\kappa_2$ and $\kappa_4$ are additive over the factors, they grow linearly with $K$. The excess kurtosis, being the ratio $\gamma_4=\kappa_4/\kappa_2^2$, therefore scales as $K/K^2=K^{-1}$. The convergence to Gaussianity is thus a direct consequence of self-averaging and does not rely on any special properties of the distributions of $\alpha_k$ and $\beta_k$ beyond the existence of the above limits.
In the final section of this paper, we shall examine in detail a particularly relevant case in which the parameters $\alpha_k$ are arbitrary, while the $\beta_k$ are independent quenched random variables drawn from an exponential distribution.
	\item   Consider finally the case in which the reduced switching rates $\alpha_k$ are confined to a finite interval, while the reduced velocities $\beta_k$ grow regularly  with $k$ according to a power law,  $\beta_k = k^p$ with $p \geq -1/2$. 
	In this case the asymptotic behavior of the excess kurtosis can be determined in a straightforward manner to give 
\begin{align}
\gamma_4 \simeq
\begin{cases}
K^{-1}, & p>-1/4,\\
(\ln K)/K, & p=-1/4,\\
K^{-(4p+2)}, & -1/2<p<-1/4,\\
(\ln K)^{-2}, & p=-1/2.
\end{cases}
\end{align}
Therefore, for $p > -1/2$ the excess kurtosis decays algebraically with the number $K$ of the dichotomous noises, with a logarithmic correction at $p=-1/4$.
On the contrary, in the marginal case $p = -1/2$ the decay is only logarithmic, reflecting a much slower approach to the Gaussian limit.
	\end{itemize}

\begin{figure}[htbp]
	\begin{center}
		\includegraphics[width=155mm]{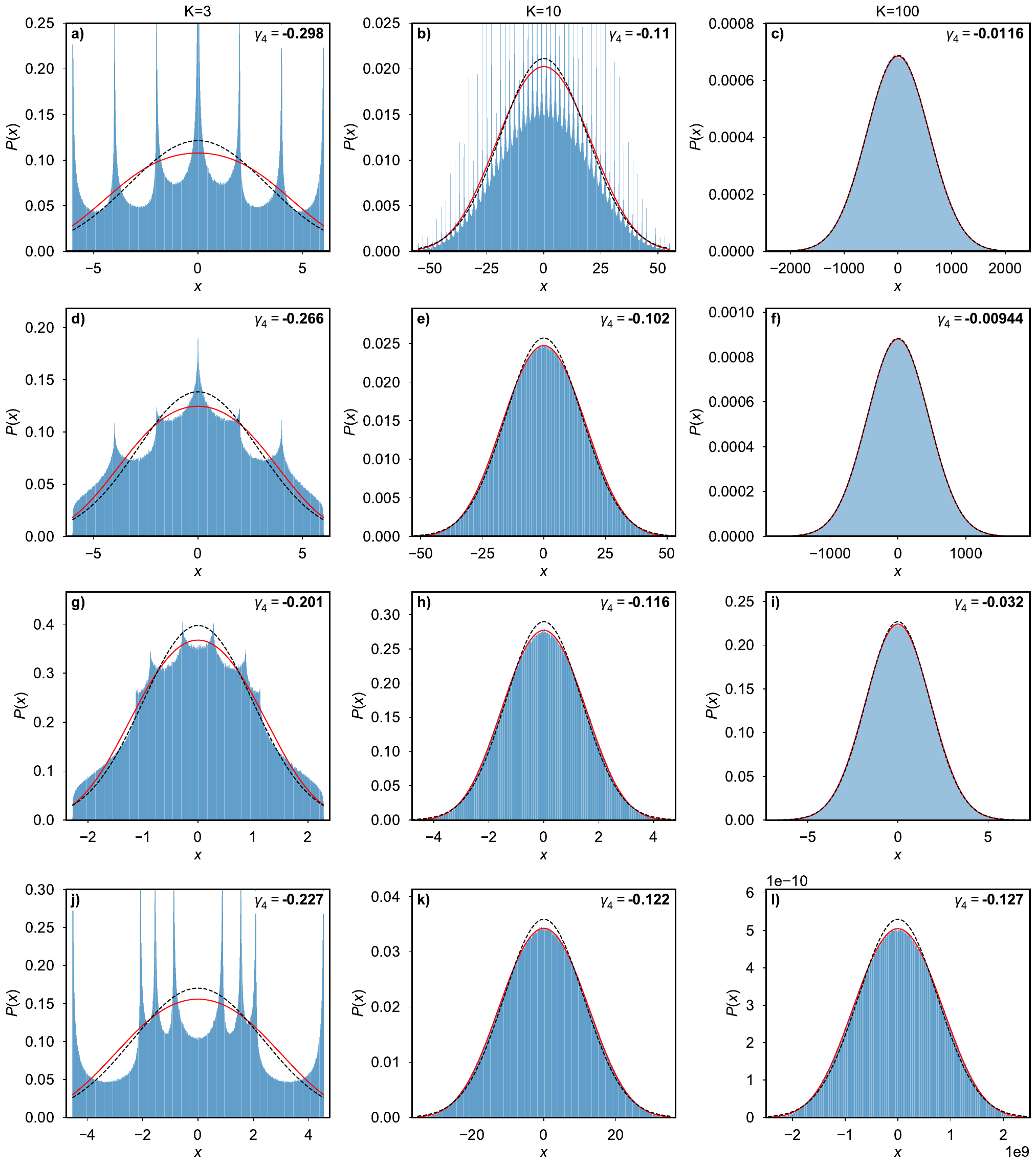}
	\end{center}
	\caption{A truncated Edgeworth series representation (see eq. \eqref{edgeworth}) of the PDF $P(x)$ for $K=3$, $10$ and $100$.  The histograms show empirical distributions obtained from numerical simulations of $10^7$ trajectories. The red solid curves show eq.~\eqref{edgeworth} including the first correction, while the black dashed curves indicate the Gaussian approximation without correction.
		In the first two rows we take $\beta_k = k$, while the values of $\alpha_k$ are taken at random from the $K$-dependent interval $(0.5/K,1/K)$ (first row) or from the interval $(0.2,0.5)$ (second row). 
		In the third and the fourth rows the values of $\alpha_k$ are taken at random from the interval $(0.2,0.5)$, while $\beta_k = 1/\sqrt{k}$ (third row) and $\beta_k = e^{k/5}$ (fourth row). }
	\label{fig:4} 
\end{figure}

Figure~\ref{fig:4} compares the stationary position PDF $P(x)$ of the process defined by eq.~\eqref{LOU}, obtained numerically for $K=3$, $10$, and $100$ independent dichotomous noises with the Gaussian approximation (the leading term in eq.~\eqref{edgeworth}, black dashed curves) and the two-term Edgeworth expansion in eq.~\eqref{edgeworth} (red solid curves). The insets display the corresponding values of the excess kurtosis $\gamma_4$.

The first two rows correspond to linearly increasing amplitudes, $\beta_k=k$ for $k=1,\ldots,K$, but differ in the choice of the reduced switching rates $\alpha_k$. In the first row, the $\alpha_k$ are drawn independently from the interval $(0.5/K,1.0/K)$, whose width decreases as $K$ increases, whereas in the second row they are drawn from the $K$-independent interval $(0.2,0.5)$. In the third and fourth rows, the reduced switching rates are again sampled from $(0.2,0.5)$, while the amplitudes follow different dependencies on $k$: in the third row, $\beta_k=1/\sqrt{k}$ decreases with $k$, whereas in the fourth row $\beta_k=e^{k/5}$ grows exponentially.

The $K=3$ (left column), illustrates that the stationary PDFs remain strongly non-Gaussian. Their support consists of several $U$-shaped segments terminating at either divergent or finite cusps, reflecting the singular structure discussed in the previous section. In these examples, the total reduced switching rate $A=\sum_{k=1}^K\alpha_k$ is still too small and the number of independent noises is insufficient for the central limit mechanism to become effective.

For $K=10$ (middle column), the PDFs have already undergone a substantial evolution toward Gaussian-like shapes. The only exception is panel (b), where the reduced switching rates are sufficiently small for the singular multi-peak structure to survive. A dense "forest" of sharp peaks located at the branch points is still evident, and these peaks carry a significant fraction of the total probability. Consequently, the stationary PDF exhibits a pronounced departure from Gaussian behavior.
In the remaining three panels, the PDFs are already close to Gaussian. Nevertheless, the leading Gaussian approximation alone still fails to reproduce the observed distributions quantitatively, whereas the inclusion of the first Edgeworth correction yields an excellent approximation over the entire range of $x$.

For $K=100$ (right column), the convergence toward Gaussianity is nearly complete in panels (c), (f), and (i): the Gaussian term alone becomes practically indistinguishable from the numerical PDF. By contrast, panel (l) demonstrates that even for such a large number of independent noises the first Edgeworth correction remains essential. This occurs because the exponentially increasing amplitudes $\beta_k=e^{k/5}$ generate a comparatively large excess kurtosis (see the inset), so that the convergence to the Gaussian limit is considerably slower.

Therefore, Fig.~\ref{fig:4} demonstrates that the emergence of Gaussian statistics is governed not only by the number $K$ of independent dichotomous noises but also by the detailed distributions of the switching rates $\alpha_k$ and amplitudes $\beta_k$. While increasing $K$ generally suppresses the non-Gaussian cumulants, the rate of this suppression depends sensitively on how the individual noises contribute to the overall fluctuations. In particular, broad or rapidly increasing amplitude spectra may preserve a significant excess kurtosis even for large $K$, making the first Edgeworth correction indispensable. This figure therefore highlights the subtle interplay between the number of independent noise sources, their characteristic switching rates, and their amplitudes in determining both the rate of convergence to the Gaussian limit and the accuracy of the Edgeworth approximation.

		\subsection{Behavior near the boundaries of the support}
\label{edges}

Consider next the behavior of the PDF in eq.~\eqref{PDF} for arbitrary values of the reduced switching rates $\alpha_k$ in the limit $x \to \pm B$. To analyse this regime, it is more convenient to abandon the formal oscillatory-integral representation in eq.~\eqref{PDF} and instead use the equivalent random-flight interpretation discussed at the end of Sec.~\ref{integral}. We find that this probabilistic picture is both technically simpler and physically more transparent. We focus in what follows on the behavior in the vicinity of the right boundary of the support; the behavior at the opposite boundary is exactly the same, by symmetry.

We therefore turn to the independent random variables $l_k$ defining the jump lengths of the random-flight shown in Fig.~\ref{fig:2}. The endpoint of the flight can lie in the vicinity of the right boundary $B$ of the support only if, for a given realization, all jump lengths $l_k$ are simultaneously close to the upper edges of their respective supports, namely to the values $\beta_k$
(see eq.~\eqref{plk}). In other words, each step must attain a value close to its maximal possible length. The behavior near the boundary of the support is therefore governed by a collective large-deviation event involving all jump variables simultaneously.

Introduce the small positive deficit variables $s_k=\beta_k-l_k\geq 0$, so that
\begin{align}
\sum_{k=1}^K l_k
=
B-\sum_{k=1}^K s_k .
\end{align}
Hence, the event in which the endpoint of the trajectory lies at a small distance $\varepsilon$ from the boundary of the support occurs when
\begin{align}
\sum_{k=1}^K s_k=\varepsilon .
\end{align}
We therefore seek the joint probability density function of the variables $s_k$ subject to the above constraint.

For small $s_k$, we have from
eq. \eqref{plk} that their 
probability density function is given approximately by
\begin{align}
	p(s_k) \approx (2 \beta_k)^{\alpha_k - 1} C_k s_k^{\alpha_k-1},
	\end{align}
	where the amplitude $C_k$ is defined in eq. \eqref{plk}. Respectively, the probability that the  sum of positive variables $s_k$ is equal to $\varepsilon$ is given by
\begin{align}
P_{\varepsilon}
= \prod_{k=1}^K (2 \beta_k)^{\alpha_k - 1} C_k 
\int_0^{\varepsilon}\ldots \int_0^{\varepsilon}
\delta\!\left(
\varepsilon-\sum_{k=1}^K s_k
\right)
\prod_{k=1}^K
s_k^{\alpha_k-1}
\, ds_k .
\end{align}
Changing the integration variables $s_k = \varepsilon \tau_k$ and using the relation $\delta(\varepsilon x) = \varepsilon^{-1} \delta(x)$, we arrive at  the following
edge asymptotics of the position PDF
\begin{align}
	P(x)
	\simeq
	 (B-x)^{A-1}  \,, \quad x \to B^{-} \,,
\end{align}
where  the superscript $"-"$ signifies that the boundary $B$ of the support is approached from below. 

A remarkable feature of the above result is that 
the behavior at the boundary of the support is very different depending whether the sum $A$ of the reduced switching rates is less,  equal or greater than $1$. If all the switching rates are sufficiently small such that their sum $A < 1$, the PDF \textit{diverges} at the boundary meaning that the trajectories of such a  random-flight, or equivalently, of the Ornstein-Uhlenbeck process driven by multiple independent dichotomous noises, accumulate near the boundary of the support. In the borderline case when $A=1$,  the PDF approaches a constant value as $x \to B^{-}$.  Conversely, for $A > 1$, the PDF vanishes at the boundary. In Fig.~\ref{fig:3} all three kinds of behavior are apparent.

\section{Random sequences $\beta_k$}
\label{random}

In this final section, we consider the situation in which the reduced switching rates $\alpha_k$ are fixed and arbitrary, while the reduced velocities $\beta_k$ are independent and identically distributed (i.i.d.) random variables drawn from a common probability distribution. In this setting, one deals not with a single stochastic process but with an ensemble of processes characterized by different realizations of the parameters $\beta_k$. For a given realization of the set ${\beta_k}$, the stationary position PDF is still given by eq.~\eqref{PDF}. However, to characterize the statistical properties of the ensemble as a whole, one must perform an additional averaging over the distribution of the random variables $\beta_k$.

Our objective is therefore to determine the disorder-averaged position probability density function $\Psi(x) = \langle P(x)\rangle_\beta$, where the brackets denote averaging with respect to the i.i.d. variables $\beta_k$. This additional level of randomness introduces quenched heterogeneity into the model, reflecting the fact that different realizations may be characterized by different velocity scales. At first sight, this setting differs from diffusing-diffusivity models, where the transport coefficient fluctuates dynamically along a given trajectory. Here, by contrast, the randomness is associated with the amplitudes $\beta_k$ of the dichotomous components and enters at the level of the ensemble of realizations rather than through temporal fluctuations of a transport parameter.

However, the connection becomes much more transparent in the equivalent random-flight representation, in which $K$ plays the role of the running time and the reduced velocities $\beta_k$ are randomly drawn at each step. In this picture, the dynamics is governed not only by the stochastic motion itself but also by an additional source of randomness in the parameters controlling that motion. This is precisely the central feature shared with diffusing-diffusivity models: the observed behavior results from a superposition of dynamical fluctuations and fluctuations of the underlying transport characteristics.
As we show below, averaging over the distribution of the reduced velocities $\beta_k$
can profoundly alter the stationary position PDF, leading to qualitative behaviors that are absent when the amplitudes are fixed.

Here we focus exclusively on the case in which the reduced velocities $\beta_k$ are i.i.d. random variables drawn from the exponential distribution
\begin{align}
	\label{betak}
	p(\beta_k)=\frac{1}{\beta}  \, e^{-\beta_k/\beta} \,,
\end{align}
where $\beta$ sets the characteristic reduced velocity scale. This choice is particularly convenient because it permits an explicit evaluation of the disorder-averaged PDF while preserving the essential effects of amplitude heterogeneity. Note also that, since the variables $\beta_k$ are supported on $[0,\infty)$, the ensemble-averaged PDF $\Psi(x) = \langle P(x)\rangle_\beta$ has now support over the entire real line, in contrast to the case considered previously.

The averaging of $P(x)$ in eq.~\eqref{PDF} can be performed straightforwardly by noting that the kernel of the Fourier integral factorizes into a product of independent contributions associated with different dichotomous components. Since the variables $\beta_k$ are i.i.d., the averaging over their distribution can be carried out separately for each factor:
\begin{align}
	\label{Znu}
	Z(\nu) =\left\langle \prod_{k=1}^{K} {\cal Z}_k(\beta_k \nu)\right\rangle_{\beta}
	= \prod_{k=1}^{K} \left\langle {\cal Z}_k(\beta_k \nu)\right\rangle_{\beta} \,,
\end{align}
where
\begin{align}
	\label{k}
	\left\langle {\cal Z}_k(\beta_k \nu)\right\rangle_{\beta}
	= \int_{0}^{\infty} d\beta_k \, p(\beta_k) \, {\cal Z}_k(\beta_k \nu) \,.
\end{align}
Consequently, the problem reduces to the calculation of a single integral for each component, after which the disorder-averaged position PDF can be reconstructed through the inverse Fourier transform. In this way, the additional averaging over the random reduced velocities preserves the overall analytical structure of the problem and allows one to investigate explicitly how disorder modifies the stationary position PDF.

Performing the integral in the right-hand-side of eq. \eqref{k}, we get
	\begin{align}
		\label{Zk}
		\begin{split}
		\langle {\cal Z}_k(\beta_k \nu)\rangle_{\beta}  
		 &= \left(1 + \beta^2 \nu^2\right)^{-1/2}\,_2F_1\left(1/2, \alpha_k - 1/2; \alpha_k + 1/2; \frac{\beta^2 \nu^2}{1 + \beta^2 \nu^2}\right) \\
		&= \,_2F_1\left(1, 1/2; \alpha_k + 1/2; - \beta^2 \nu^2\right)\,,
		\end{split}
\end{align}
where ${}_2F_1(\ldots)$ is the Gauss hypergeometric function \cite{bateman}. In the analysis that follows, we will make use of either representation of $\langle {\cal Z}_k(\beta_k \nu)\rangle_{\beta}$ given in the first or in the second lines above, depending on which form is more convenient for the particular calculation.

A few remarks concerning the behavior of ${}_2F_1(\ldots)$ in the first line in the above expression are in order. As one may readily check,  this function is analytic and bounded for all $\nu \in [0,\infty)$, varying monotonically between its value $1$ at $\nu=0$, and the limiting value $\sqrt{\pi} \, \Gamma(\alpha_k + 1/2)/\Gamma(\alpha_k)$
as $\nu\to\infty$.  For $\alpha_k > 1/2$ it is a monotonically \textit{increasing} function because $\sqrt{\pi} \, \Gamma(\alpha_k + 1/2)/\Gamma(\alpha_k) > 1$. Conversely, for $\alpha_k < 1/2$ it is a monotonically \textit{decreasing} function of $\nu$. In the borderline case $\alpha_k = 1/2$ the above hypergeometric function is identically equal to $1$, such that each factor with $\alpha_k = 1/2$ contributes only $1/\sqrt{1 + \beta^2 \nu^2}$ to the kernel.

Consequently, the large-$\nu$ behavior of $\langle {\cal Z}_k(\beta_k \nu)\rangle_{\beta}$ and hence, of $Z(\nu)$ defined in eq. \eqref{Znu}, is entirely defined by the factor $1/\sqrt{1 + \beta^2 \nu^2}$ in the right-hand-side of eq. \eqref{Zk}. This implies, in particular, that the disorder-averaged position PDF,
\begin{align}
	\label{Psi}
	\Psi(x) = \langle P(x) \rangle_{\beta} = \frac{1}{\pi} \int^{\infty}_0 d\nu \, \cos(x \nu) \, Z(\nu) \,,
\end{align}
diverges at $x = 0$ for $K=1$, and stays finite for any $K > 1$. Also for finite $x$ the integral in the above expression converges rapidly when $K$ becomes larger than $1$.

\subsection{General formula for $\Psi(x)$}

The formally exact expression for $\Psi(x)$, for arbitrary $x$ and $K$, can be evaluated using the following approach: We use first the series representation  \cite{bateman}
\begin{align}
	\begin{split}
&{}_2F_1\left(1/2, \alpha_k - 1/2; \alpha_k + 1/2; \frac{\beta^2 \nu^2}{1 + \beta^2 \nu^2}\right)	= \frac{\Gamma(\alpha_k +1/2)}{\sqrt{\pi} \, \Gamma(\alpha_k - 1/2)} \\ &\times \sum_{n=0}^{\infty}\frac{\Gamma\left(n+1/2\right)
	\Gamma\left(n+a_k-1/2\right)}{\Gamma\left(n+a_k+1/2\right) n!} \, \left(\frac{\beta^2 \nu^2}{1 + \beta^2 \nu^2}\right)^n
	\end{split}
	\end{align}
to get
\begin{align}
	\begin{split}
	\langle {\cal Z}_k(\beta_k \nu)\rangle_{\beta} &=  
\frac{\Gamma(\alpha_k +1/2)}{\sqrt{\pi} \, \Gamma(\alpha_k - 1/2) \, \beta}  \sum_{n=0}^{\infty} f_k(n) \, \frac{\nu^{2 n}}{\left(1/\beta^2 +  \nu^2\right)^{n+1/2}} \\
f_k(n) &= \frac{\Gamma\left(n+1/2\right)
	\Gamma\left(n+a_k-1/2\right)}{\Gamma\left(n+a_k+1/2\right) n!} \,.
	\end{split}
	\end{align}
Using next the standard formula for the series expansion of the product of series
\begin{align}
	g(z) = \prod_{k=1}^K \left(\sum_{n=0}^{\infty} f_k(n) z^n\right) = \sum_{n=0}^{\infty} g(n) z^n \,,
	\end{align}
where the coefficients $g(n)$ are the $K$-fold discrete convolutions of the sequences $f_1(n), f_2(n), \ldots, f_K(n)$; that being, 
\begin{align}
	g(n) = \sum_{m_1 + m_2 + \ldots m_K = n}  \prod_{k=1}^K f_k(m_k) \,, 
	\end{align}
	with the sum extending over all partitions of $n$ into $K$ non-negative integer parts $m_k$, we find that the product $Z(\nu)$ in eq. \eqref{Znu} obeys
\begin{align}
	\begin{split}
		Z(\nu) &= \left(\prod_{k=1}^K \frac{\Gamma(\alpha_k +1/2)}{\Gamma(\alpha_k -1/2)} \right) \left(\sqrt{\pi} \beta \right)^{-K} \\
		& \sum_{n=0}^{\infty} \left(\sum_{m_1 + m_2 + \ldots m_K = n}  \prod_{k=1}^K f_k(m_k)\right) \frac{\nu^{2 n}}{\left(1/\beta^2 +  \nu^2\right)^{n+K/2}} \,.
		\end{split}
	\end{align}
Upon inserting the above expression into eq.~\eqref{Psi}, the problem reduces to the evaluation of the integral
\begin{align}
	\begin{split}
	\int^{\infty}_0 \frac{\nu^{2 n} \, \cos(x \nu) \, d\nu}{\left(1/\beta^2 + \nu^2\right)^{n + K/2}} &= (-1)^n \frac{\sqrt{\pi}}{\Gamma(n+ K/2)} \left(\frac{\beta}{2}\right)^{n + (K-1)/2} \\
	&\times \frac{d^{2n}}{d x^{2n}} \left[|x|^{n+(K-1)/2}   K_{n+(K-1)/2}\left(\frac{|x|}{\beta}\right)\right] \,,
	\end{split}
	\end{align}
	where $K_r(\ldots)$ is the modified Bessel function of the Second kind \cite{WATSON}. 
Consequently, the desired disorder-averaged PDF $\Psi(x)$ is given by
\begin{align}
	\begin{split}
		\label{gen}
		&\Psi(x) = \frac{2}{(2 \pi \beta)^{(K+1)/2}} \left(\prod_{k=1}^K \frac{\Gamma(\alpha_k +1/2)}{\Gamma(\alpha_k -1/2)} \right)  \sum_{n=0}^{\infty} \frac{(-1)^n}{\Gamma(n + K/2)} \\
		&\times \left(\sum_{m_1 + m_2 + \ldots m_K = n}  \prod_{k=1}^K f_k(m_k)\right) \left(\frac{\beta}{2}\right)^{n} 
		\frac{d^{2n}}{d x^{2n}} \left[|x|^{n+(K-1)/2}   K_{n+(K-1)/2}\left(\frac{|x|}{\beta}\right)\right] \,,
		\end{split}
		\end{align}
		which is formally valid for arbitrary sequence of the reduced switching rates $\alpha_k$ and arbitrary $x$.

\subsection{Asymptotic large-$x$ behavior of $\Psi(x)$}

Since $K_r(|x|/\beta) \sim \exp(-|x|/\beta)$ as $|x| \to \infty$ regardless of the value of $r$ \cite{WATSON}, it is immediately apparent from eq. \eqref{gen} that the leading large-$|x|$ behavior of $\Psi(x)$ is exponential, $\Psi(x) \sim \exp(-|x|/\beta)$,
likewise the parental distribution in eq. \eqref{betak} of amplitudes $\beta_k$.
While the displacement distribution has compact support for any fixed realization of the amplitudes $\beta_k$, averaging over these amplitudes generates exponential tails, in qualitative agreement with diffusing-diffusivity models. The full asymptotic behavior may be, however, a bit more complicated. In general, the exponential factor may be multiplied by an algebraic function of $|x|$ whose exponent depends in a nontrivial way on the set of fixed $\alpha_k$. This information is not easily accessible from the representation in eq. \eqref{gen}. 
To uncover it, we revisit the Fourier representation \eqref{Psi} and perform a singularity analysis  of the kernel in the complex $\nu$-plane; namely, in the vicinity of the points $\nu = \pm i/\beta$. 
The nature of the singularity then determines the precise asymptotic form of $\Psi(x)$, including both the exponential decay and its power-law prefactor.

Some key features of the general model are already present in the simplest case $K=1$, which we consider first. In this case, the integral in eq.~\eqref{Psi} can be evaluated exactly (see \cite{prud}) for arbitrary $\alpha_1$ and $\beta$. The resulting expression, however, is not particularly illuminating: it involves a combination of two generalized hypergeometric functions $_3F_4(\ldots)$ \cite{bateman}, which diverge (with opposite signs) for the parameter values relevant to our model. Upon regularization, one obtains a rather cumbersome series expansion in powers of $x$, whose coefficients involve polylogarithmic functions, making a direct analysis of the short- and large-$x$ behavior rather difficult. Mathematica is also able to perform the integral, but returns the result in terms of a Meijer G-function, which is likewise not especially convenient for extracting asymptotic properties. 

\begin{figure}[t]
	\begin{center}
		\includegraphics[width=120mm]{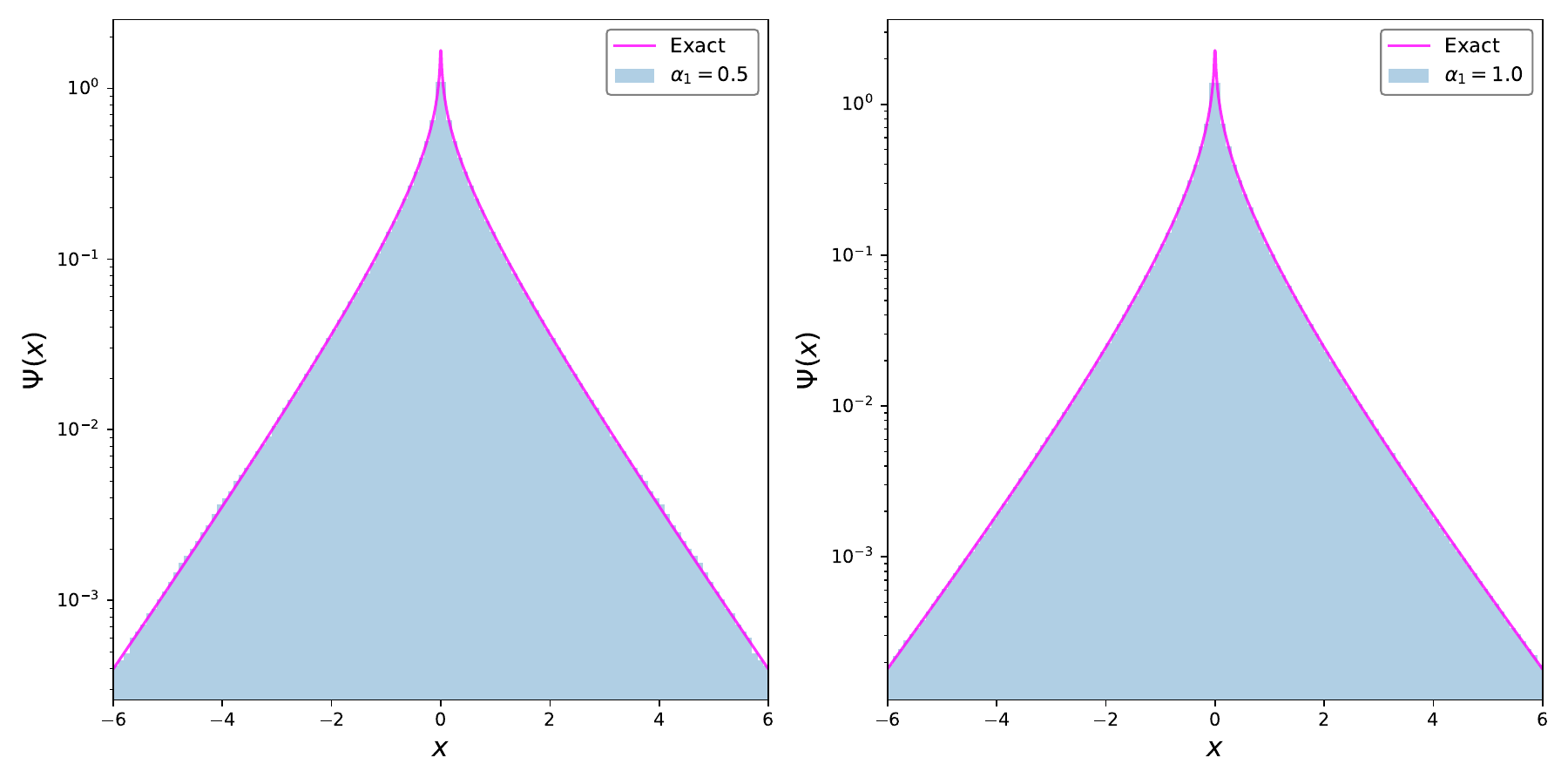}
	\end{center}
	\caption{The disorder-averaged position PDF $\Psi(x)$ for $K = 1$ and $\beta = 1$. Panel (a):  $\alpha_1=1/2$.  Panel (b): $\alpha_1=1$.
	Thin magenta curves depict the exact analytical solutions presented in the text, while the blue histograms show the results of numerical simulations for the corresponding values of $\alpha_1$.}
	\label{fig:5} 
	\end{figure}

Significant simplifications occur for particular values of $\alpha_1$. For instance, one finds 
$\Psi(x) = K_0(|x|/\beta)/(\pi \beta)$ for $\alpha_1=1/2$,
which result coincides (rather incidentally) with 
the short-time displacement distribution of the standard diffusing-diffusivity model \cite{chechkin2017,denis}. For $\alpha_1 = 1$, one obtains $\Psi(x)  = - Ei(-|x|/\beta)/(2 \beta)$, where $Ei(z)$ is the exponential integral \cite{bateman}. 
Both expressions are drawn in Fig.~\ref{fig:5} and compared against the results of numerical simulations demonstrating an excellent agreement between analytical predictions and numerical data. 

The salient feature of the above expressions is that they both diverge logarithmically as $x \to 0$, in agreement with our general prediction for the case $K=1$.  Concurrently, both decay exponentially as $|x| \to \infty$. More specifically, for $\alpha_1 = 1/2$ we have $\Psi(x) \simeq \exp(-|x|/\beta)/\sqrt{|x|}$, while for $\alpha_1 = 1$ we get $\Psi(x) \simeq \exp(-|x|/\beta)/|x|$.
These observations suggest that, for arbitrary $\alpha_1$, the disorder-averaged position PDF should exhibit the asymptotic behavior $\Psi(x) \simeq \exp(-|x|/\beta)/|x|^{\alpha_1}$ as $|x| \to \infty$. 
To show that this is indeed the case for arbitrary values of $\alpha_1$, it is expedient to use the integral representation of the Gauss hypergeometric function \cite{bateman} 
\begin{align}
	\label{F}
	\,_2F_1\left(1, 1/2; \alpha_k + 1/2; - \beta^2 \nu^2\right) = \frac{\Gamma(\alpha_k + 1/2)}{\sqrt{\pi} \, \Gamma(\alpha_k)} \int^1_0 \frac{\left(1 - \xi\right)^{\alpha_1 - 1} \, d\xi}{\sqrt{\xi} \, \left(1 + \beta^2 \nu^2 \xi\right)}
\end{align}
and perform first the integration over the variable $\nu$. This yields the following exact integral representation of the disorder-averaged position PDF for $K=1$:
\begin{align}
	\Psi(x) = \frac{\Gamma(\alpha_1 + 1/2)}{2 \sqrt{\pi} \,\Gamma(\alpha_1)\,\beta}
	\int_0^1 \frac{d\xi}{\xi \,(1-\xi)^{1-\alpha_1}}
	\exp\left(-\frac{|x|}{\beta\sqrt{\xi}}\right).
\end{align}
Two important conclusions follow immediately from this expression. First, independently of the value of $\alpha_1$, the integral develops a logarithmic divergence as $x \to 0$, showing that the singularity at the origin is indeed a generic feature of the disorder-averaged distribution in the case $K=1$.
Second, in the limit 
$|x| \to \infty$, 
the dominant contribution arises from the vicinity of the upper integration limit $\xi = 1$. A standard Laplace analysis then gives the following asymptotic expansion
\begin{align}
	\Psi(x) =
	\frac{\Gamma(\alpha_1 + 1/2)}{\sqrt{\pi}}
	\frac{(2 \beta)^{\alpha_1-1}}{|x|^{\alpha_1}}
	e^{-|x|/\beta}
	\left[
	1+\frac{\alpha_1(1-3\alpha_1)\beta}{2|x|}
	+ O\left(\frac{1}{x^2}\right)
	\right].
\end{align}
This result confirms the conjectured asymptotic form and shows that the algebraic prefactor is entirely controlled by the reduced switching rate $\alpha_1$, while the exponential decay length is set by the characteristic amplitude scale $\beta$. 

Consider next the case $K=2$, in which a more non-trivial behavior emerges. We consider two situations: (i) $\alpha_1$ and $\alpha_2$ are both strictly less than $1$; and (ii) $\alpha_1 \geq 1$ and $\alpha_2 > 1$.
We state here the final results and defer the technical derivations to \ref{B}.
We find that in these two cases the leading asymptotic behavior of the disorder-averaged position PDF in the limit $|x| \to \infty$ is given by 
\begin{align}
	\label{k2}
\Psi(x) \simeq \frac{\exp\left(-|x|/\beta\right)}{|x|^{\mu}} \,, \quad \mu	= \begin{cases}
	\alpha_1 + \alpha_2 - 1 \,, \qquad \qquad \text{case (i)} \,, \\
	\min(\alpha_1,\alpha_2) \,, \, \qquad \qquad \text{case (ii)}  \,.
\end{cases}
\end{align}

\begin{figure}[htbp]
	\begin{center}
		\includegraphics[width=120mm]{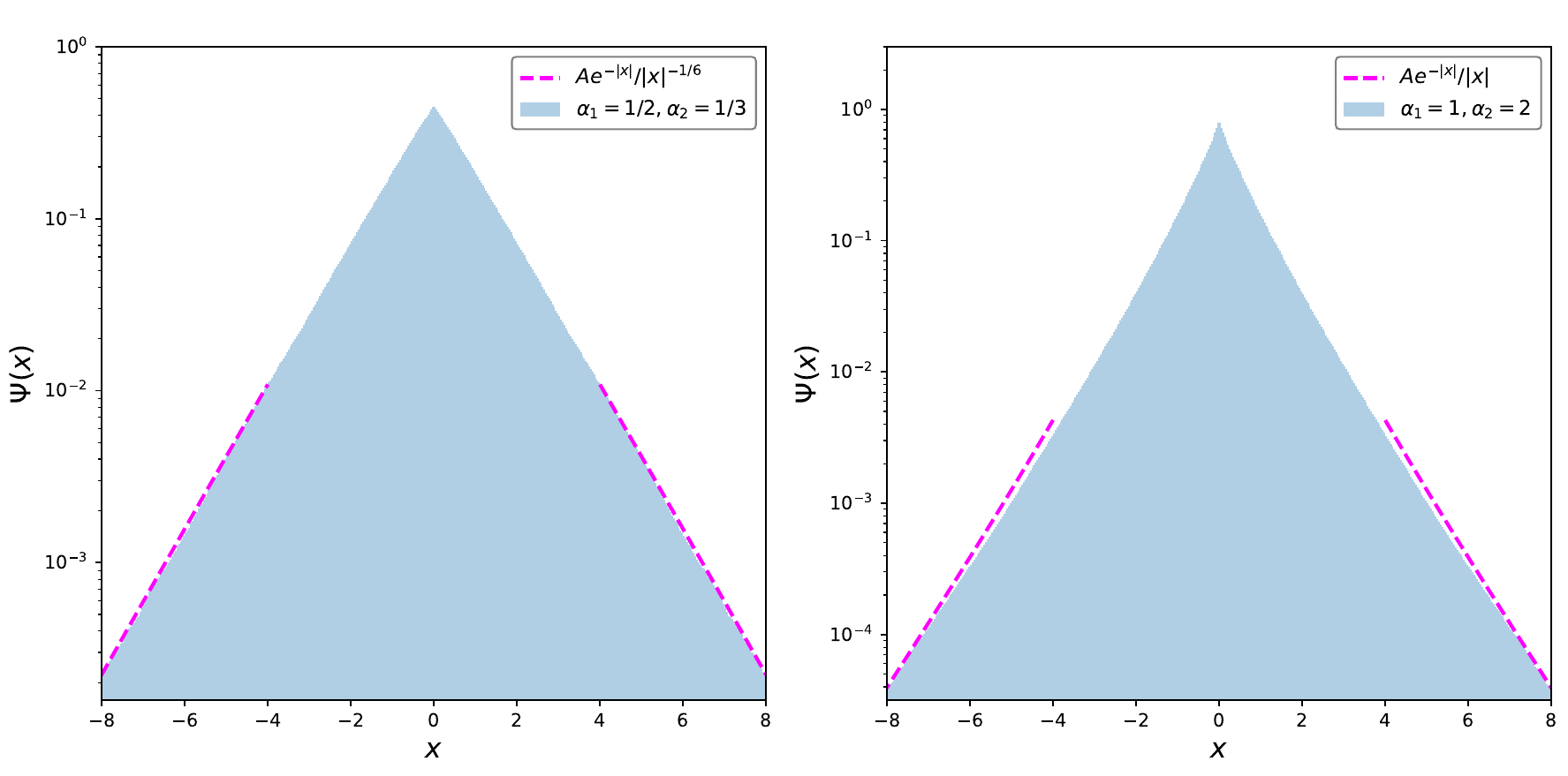}
	\end{center}
	\caption{The disorder-averaged position PDF $\Psi(x)$ for $K=2$ and $\beta=1$. In the left panel, we take $\alpha_1=1/2$ and $\alpha_2=1/3$, while in the right panel, we take $\alpha_1=1$ and $\alpha_2=2$. These correspond to cases (i) and (ii), with $\mu=-1/6$ and $\mu=1$, respectively. Dashed magenta curves show the asymptotic predictions in eq. \eqref{k2}, while the histograms show the corresponding numerical simulation results.
	}
	\label{fig:6} 
\end{figure}

The above result reveals two remarkable features of the disorder-averaged position PDF. When $\alpha_1,\alpha_2 < 1$, the large-$|x|$ asymptotics is governed by the combined action of both noises, with the exponent $\mu = \alpha_1 + \alpha_2 - 1$.
Consequently, $\mu > 0$ when $\alpha_1 + \alpha_2 > 1$, so that the algebraic prefactor enhances the leading exponential decay. At the marginal point $\alpha_1 + \alpha_2 = 1$, the algebraic correction disappears, whereas for $\alpha_1 + \alpha_2 < 1$ one has $\mu < 0$, leading to a slower decay than the pure exponential law.
This "cooperative" behavior disappears in the case (ii). In this case, the tail of $\Psi(x)$ is controlled entirely by the smaller $\alpha_k$, while the contribution of the second noise becomes asymptotically negligible. This outcome is far from obvious, given that the two noises enter the Langevin equation \eqref{LOU} on an equal footing through an additive forcing. The large-$|x|$ behavior therefore exhibits a form of asymptotic selection: despite the presence of two noise sources, the tail of the PDF is ultimately governed by a single component with the smaller $\alpha_k$.
Finally, we note that we have deliberately excluded from our analysis, (as well as from the subsequent analysis for arbitrary  $K \geq 2$ presented below), 
the marginal cases when $\alpha_1 = 1$, while $\alpha_2 \leq 1$, in which the large-$|x|$ behavior
acquires additional logarithmic corrections (see eq.~\eqref{log} in \ref{B}). Note that, on contrary, the  asymptotic  behavior of $\Psi(x)$ in the case $\alpha_1=1$ and $\alpha_2 > 1$ obeys our eq. \eqref{k2}.

Figure~\ref{fig:6} shows the disorder-averaged position PDF $\Psi(x)$ for $K=2$. 
In the left panel, we take $\alpha_1=1/2$ and $\alpha_2=1/3$, and in the right panel, $\alpha_1=1$ and $\alpha_2=2$. 
These correspond to cases (i) and (ii), with the power-law correction with exponent $\mu=-1/6$ and $\mu=1$, respectively. 
In both cases, we observe good agreement between the numerical simulations and the predictions of the asymptotic tail behaviors in eq. \eqref{k2}.
For a direct verification of the predicted power-law corrections, see Fig.~\ref{fig:S1}.

We now turn to the general case of arbitrary $K$. To ease the discussion while still capturing the essential features, we focus on two representative situations: \\
(i) $m$ of the reduced switching rates $\alpha_k$, (with $1 \leq  m \leq K$), are smaller than unity, while the remaining $K-m$ rates are all greater than $1$,\\
(ii) either $\alpha_k > 1$ for all $k=1, 2, \ldots, K$, or there exists a unique index $j$ such that $\alpha_j=1$ and $\alpha_k > 1$ for all $k \neq j$.

Relegating the technical details of the derivation to \ref{B}, we find that the disorder-averaged position PDF in eq.~\eqref{Psi} exhibits the following large-$|x|$ asymptotic behavior:
\begin{align}
	\label{K}
	\Psi(x) \simeq \frac{\exp\left(-|x|/\beta\right)}{|x|^{\mu}} \,, \quad \mu	= \begin{cases}
		\sum_{\alpha_k < 1} \alpha_k - m + 1 \,, \qquad \quad \,\,\, \text{case (i)} \,, \\
		\min(\alpha_1, \alpha_2, \ldots, \alpha_K) \,, \qquad \quad \text{case (ii)} \,,\\
	\end{cases}
\end{align}
where the summation symbol with the subscript $\alpha_k < 1$ signifies that the summation extends over the values of $\alpha_k$ which are less than $1$.

\begin{figure}[t]
	\begin{center}
		\includegraphics[width=155mm]{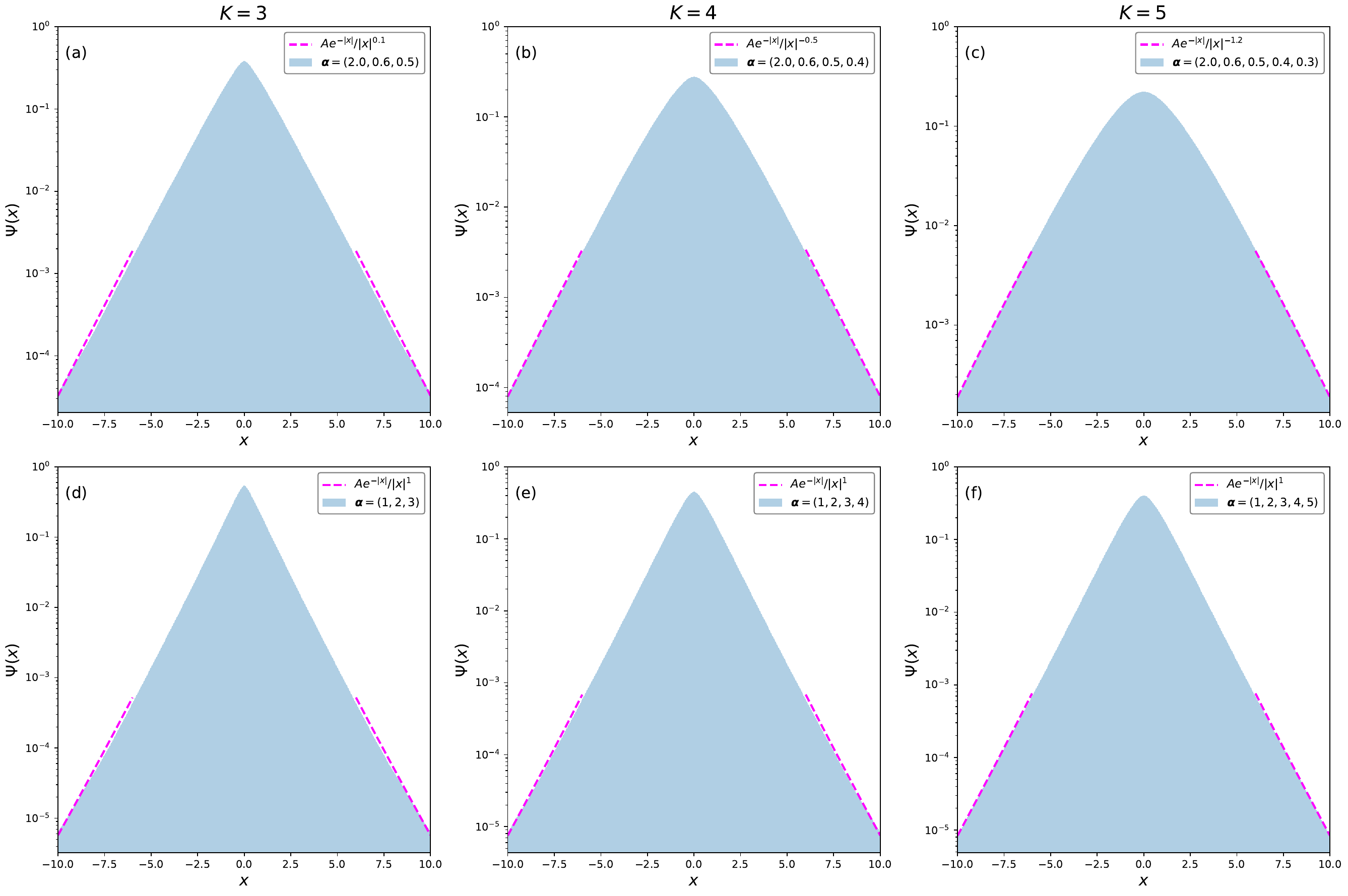}
	\end{center}
	\caption{The disorder-averaged position PDF $\Psi(x)$ for $K=3$, $4$, and $5$, with $\beta=1$. The first row corresponds to case (i) of eq.~\eqref{K}, with the reduced switching rates $\boldsymbol{\alpha}=(2.0, 0.6, 0.5)$, $(2.0, 0.6, 0.5, 0.4)$, and $(2.0, 0.6, 0.5, 0.4, 0.3)$ for $K=3$, $4$, and $5$, respectively. The second row corresponds to case (ii), with $\boldsymbol{\alpha}=(1.0, 2.0, 3.0)$, $(1.0, 2.0, 3.0, 4.0)$, and $(1.0, 2.0, 3.0, 4.0, 5.0)$ for $K=3$, $4$, and $5$, respectively. Dashed magenta curves show the asymptotic predictions of eq.~\eqref{K}, while the histograms show the numerical simulation results.
	}
	\label{fig:7} 
\end{figure}

This result generalizes in a simple way the behavior obtained previously for $K=2$. In case (i), the exponent of the power-law prefactor is determined solely by the $m$ reduced switching rates that are smaller than unity. Each such rate contributes additively to the exponent, whereas all rates exceeding unity become asymptotically irrelevant - they affect only the amplitude and leave the functional form of the decay unchanged. In this sense, the slowly switching components completely control the right-tail behavior of the disorder-averaged PDF. 
Note that the exponent $\mu = \sum_{\alpha_k < 1} \alpha_k - m + 1$ can be positive for some choices of $\alpha_k$, in which case the algebraic prefactor will accelerate the decay of $\Psi(x)$, can be equal to zero when the sum $\sum_{\alpha_k < 1} \alpha_k = m - 1$, or can be negative, slowing down the decay.
Note, as well, that for $K = 2$ and $m=2$ the result in the first line in eq.~\eqref{K} reduces to the result presented in the first line in eq. \eqref{k2}. 

Conversely, in case (ii), in which all reduced switching rates exceed unity, the asymptotic decay is governed entirely by the smallest rate among them. 
If we permit one of $\alpha_k$ be equal to $1$, then 
we will evidently have $\mu = 1$. Remarkably, despite the additive nature of the underlying noises, the contribution of all larger $\alpha_k$ is asymptotically suppressed, and the tail of $\Psi(x)$
is controlled by a single dominant mode corresponding to the slowest switching process. This is also consistent with our prediction in eq. \eqref{k2}.

Figure~\ref{fig:7} shows the disorder-averaged position PDF $\Psi(x)$ for $K=3$, $4$, and $5$.
In the first row, we take $\boldsymbol{\alpha}=(2.0,0.6,0.5)$, $(2.0,0.6,0.5,0.4)$, and $(2.0,0.6,0.5,0.4,0.3)$ for $K=3$, $4$, and $5$, respectively.
These parameter sets correspond to case (i) of eq.~\eqref{K}, with $\mu=0.1$, $-0.5$, and $-1.2$, respectively.
In the second row, we take $\boldsymbol{\alpha}=(1.0,2.0,3.0)$, $(1.0,2.0,3.0,4.0)$, and $(1.0,2.0,3.0,4.0,5.0)$ for $K=3$, $4$, and $5$, respectively, corresponding to case (ii).
Since $\alpha_1=1$ is the smallest reduced switching rate in all three cases, $\mu=1$ for all three values of $K$.
The dashed magenta curves indicate the asymptotic predictions of eq.~\eqref{K}, while the histograms show the numerical simulation results.
In all cases, the asymptotic predictions are consistent with the large-$|x|$ behavior observed in the simulations.
For a direct verification of the predicted power-law corrections, see Fig.~\ref{fig:S2}.

\section{Conclusions}
\label{concs}

In this work, we have introduced and solved a generalized Ornstein--Uhlenbeck process driven by a finite superposition of independent dichotomous noises with arbitrary amplitudes and switching rates. We derived two exact representations of the stationary position probability density function (PDF): an oscillatory integral involving products of Bessel functions and an equivalent Fourier-series expansion. These complementary representations allowed us to characterize the stationary measure in detail. We showed that it possesses compact support and, in general, exhibits a hierarchy of algebraic singularities located at all signed combinations of the characteristic noise amplitudes. Depending on the choice of parameters, the stationary PDF may display several local extrema as well as cusp and edge singularities. We also interpreted the model in terms of an equivalent finite-step random-flight process with bounded independent jumps. This probabilistic picture provides an intuitive understanding of the stationary distribution and clarifies the origin of its singularities.

We also analyzed the stationary PDF through its complete hierarchy of cumulants and the corresponding Edgeworth expansion. These results allowed us to identify the conditions under which the stationary distribution converges to a Gaussian and to quantify the leading deviations from Gaussianity whenever these conditions are not fully satisfied.

The framework also admits a natural extension to quenched disorder. For exponentially distributed reduced noise amplitudes and fixed switching rates, we obtained an exact representation of the disorder-averaged stationary PDF. In contrast to the compactly supported distributions of the deterministic model, the disorder average generates exponentially decaying tails multiplied by algebraic prefactors whose exponents depend on the switching rates. We further showed that the asymptotic behavior is governed by a competition between different noise components, leading either to collective contributions from several sources or to the dominance of a single component.

Beyond the exact solution of the model itself, the present work identifies a class of colored-noise processes that remains analytically tractable despite being intrinsically non-Gaussian. The results demonstrate that a superposition of elementary finite-state stochastic processes is sufficient to generate stationary distributions with nontrivial singular structures, compact support, and a controlled crossover to Gaussian behavior. The exact expressions obtained here therefore provide a useful reference for the analysis of stochastic systems driven by finite-state colored noises.

Several directions appear worth pursuing. 
A natural next step is to investigate the relaxation dynamics of the present model, in particular its approach toward the stationary state characterized here. Further extensions include nonlinear confining potentials, multiplicative and interacting dichotomous noises, time-dependent forcing, and higher-dimensional generalizations. 
The model may also serve as a minimal framework for describing systems evolving in switching environments, including problems in active and biological matter, stochastic transport, gene-regulatory dynamics, and regime-switching models in finance, where fluctuations originate from nonequilibrium environments with a finite number of internal states.

\section*{Acknowledgments}

The authors acknowledge helpful discussions with S. Majumdar and L.-H. Tang.  The National Research
Foundation of Korea, Grants No. RS-2024-00343900, is acknowledged (J.-H.J).

\appendix

\section{Cumulants of the position probability density function}
		\label{C}

We start with the definition of the cumulant generating function
\begin{equation}
	\label{eq:defcgf}
H(\nu) =	\log{\Phi\left(\nu\right)}=\sum_{n=1}^{\infty}\kappa_n\frac{\left(i\nu\right)^n}{n!}=\sum_{n=1}^{\infty}\kappa_{2n}\frac{\left(-1\right)^n\nu^{2n}}{\left(2n\right)!} \,,
\end{equation}
where $\Phi(\nu)$ is the characteristic function defined in eqs. \eqref{Phi} and \eqref{Z}, while $\kappa_{2 n}$ denote the desired cumulants of the even order. The cumulants of odd order vanish due to symmetry. 

Next, the key step is to  use the Weierstrass representation of the Bessel function of order $\alpha_k-1/2$ in form of an infinite product:
\begin{equation}
	J_{\alpha_k-1/2}\left(\beta_k \nu\right)=\frac{1}{\Gamma\left(\alpha_k+\frac{1}{2}\right)}\left(\frac{\beta_k \nu}{2}\right)^{\alpha_k-\frac{1}{2}}\prod_{m=1}^{\infty}\left(1-\frac{\beta_k^2\nu^2}{j^2_{\alpha_k-1/2,\,m}}\right),
\end{equation}
where $j_{\alpha_k - 1/2,m}$ is the $m$-th zero (organized in the ascending order) of $J_{\alpha_k - 1/2}\left(z\right)$. This gives the following representation of $\mathcal{Z}_k\left(\beta_k \nu\right)$ in eq. \eqref{Z}
	\begin{equation}
	\mathcal{Z}_k\left(\beta_k \nu\right)=\prod_{m=1}^{\infty}\left(1-\frac{\beta_k^2\nu^2}{j^2_{\alpha_k-1/2,\,m}}\right),
\end{equation}
and hence, its logarithm obeys
\begin{align}
	\log{\mathcal{Z}_k\left(\beta_k \nu\right)}&=\sum_{m=1}^{\infty}\log{\left(1-\frac{\beta_k^2\nu^2}{j^2_{\alpha_k-1/2,\,m}}\right)}\\
	&=-\sum_{n=1}^{\infty}\left(\frac{\beta_k^{2n}}{n}\sum_{m=1}^{\infty}\frac{1}{j^{2n}_{\alpha_k-\frac{1}{2},\,m}}\right)\nu^{2n}\\
	&=-\sum_{n=1}^{\infty}\frac{\beta_{k}^{2n}\,\sigma_{2n}\left(\alpha_k-\frac{1}{2}\right)}{n}\nu^{2n}
\end{align}
where $\sigma_{2n}\left(\alpha_k-\frac{1}{2}\right)$ is  the Rayleigh function \cite{kishore}
\begin{equation}
	\label{rayleigh}
	\sigma_{2n}\left(\alpha_k-\frac{1}{2}\right)=\sum_{m=1}^{\infty}\frac{1}{j^{2n}_{\alpha_k-\frac{1}{2},\,m}}. 
\end{equation}
Lastly, inserting the above expansion into the definition of the cumulant generating function,
\begin{equation}
	H(\nu)= \log{\Phi\left(\nu\right)}=\sum_{k=1}^K\log{\mathcal{Z}_k\left(\beta_k\nu\right)},
\end{equation}
and merely comparing it with the expansion in eq. \eqref{eq:defcgf}, we readily deduce the result for the cumulants of arbitrary even order presented in eq. \eqref{kappan} in the main text.

\section{Large-$x$ asymptotics of $\Psi(x)$ in eq. \eqref{Psi}}
\label{B}

Consider first the case $K = 2$. Using the integral representation of the Gauss hypergeometric function in eq. \eqref{F} we find that $\Psi(x)$ is formally given by
\begin{align}
	\begin{split}
		\label{inn}
		\Psi(x) &= \frac{\Gamma(\alpha_1 + 1/2) \Gamma(\alpha_2+ 1/2)}{\pi^2 \Gamma(\alpha_1) \Gamma(\alpha_2)} \int^{\infty}_0 d\nu  \int^1_0 \int^1_0  \frac{(1-\xi_1)^{\alpha_1 - 1} (1-\xi_2)^{\alpha_2 - 1} \, \cos(\nu x) d\xi_1 d\xi_2}{(\xi_1 \xi_2)^{1/2} (1 + \beta^2 \nu^2 \xi_1) (1 + \beta^2 \nu^2 \xi_2)} \\
		&= \frac{\Gamma(\alpha_1 + 1/2) \Gamma(\alpha_2+ 1/2)}{\pi^2 \Gamma(\alpha_1) \Gamma(\alpha_2)} \int^1_0 \int^1_0  \frac{(1-\xi_1)^{\alpha_1 - 1} (1-\xi_2)^{\alpha_2 - 1} \, d\xi_1 d\xi_2}{(\xi_1 \xi_2)^{1/2}} \\
		& \times \int^{\infty}_0 d\nu \, \Bigg[\frac{\xi_1 \, \cos(\nu x)}{(\xi_1 - \xi_2) (1 + \beta^2 \nu^2 \xi_1)}+ \frac{\xi_2 \, \cos(\nu x)}{(\xi_2 - \xi_1) (1 + \beta^2 \nu^2 \xi_2)}\Bigg] \,.
	\end{split}
\end{align}
Performing the integral over $d\nu$, we get
\begin{align}
	\begin{split}
		\label{in1}
		\Psi(x) &= \frac{\Gamma(\alpha_1 + 1/2) \Gamma(\alpha_2+ 1/2)}{2 \pi \Gamma(\alpha_1) \Gamma(\alpha_2) \beta} \int^1_0 \int^1_0  \frac{(1-\xi_1)^{\alpha_1 - 1} (1-\xi_2)^{\alpha_2 - 1} \, d\xi_1 d\xi_2}{(\xi_1 \xi_2)^{1/2}} \\
		&\times \Bigg[\frac{\sqrt{\xi_1}}{\xi_1 - \xi_2} \exp\left(-\frac{|x|}{\beta \sqrt{\xi_1}}\right) + \frac{\sqrt{\xi_2}}{\xi_2 - \xi_1} \exp\left(-\frac{|x|}{\beta \sqrt{\xi_2}}\right)\Bigg] \,,
	\end{split}
\end{align}
which therefore naturally splits into two contributions: \\
In the first two-fold integral $I_1(x)$ the condition $|x| \to \infty$ implies that $\xi_1$ is localized in the vicinity of  $\xi_1 = 1$, while the location of the support of integration over $d\xi_2$ depends on the value of $\alpha_2$. For $\alpha_2 > 1$, the support of the integral is in the bulk of the interval $(0,1)$, away from the boundary $\xi_2 = 1$. Conversely, for $\alpha_2 < 1$, the dominant contribution to the integral over $\xi_2$ comes from the narrow  boundary layer $1 - \xi_2 = O(1/x)$. 
In other words, in this latter case both variables are localized near the corner $\xi_1 \approx 1$ and $\xi_2 \approx 1$. Performing then a standard asymptotic analysis, we find that the first two-fold integral vanishes as
\begin{align}
	\label{I1}
	I_1(x) \simeq \frac{\exp\left(-|x|/\beta\right)}{|x|^{\mu_1}} \,, \quad \mu_1	= \begin{cases}
		\alpha_1 \,, \qquad \qquad \,\,\,\, \alpha_2 > 1 \\
		\alpha_1 + \alpha_2 - 1 \,, \quad \alpha_1, \alpha_2 < 1 \,.
	\end{cases}
\end{align}
In a similar way we analyze the contribution of the second two-fold integral to get that, by symmetry,
\begin{align}
	\label{I2}
	I_2(x) \simeq \frac{\exp\left(-|x|/\beta\right)}{|x|^{\mu_2}} \,, \quad \mu_2	= \begin{cases}
		\alpha_2 \,, \qquad \qquad \,\,\,\, \alpha_1 > 1 \\
		\alpha_1 + \alpha_2 - 1 \,, \quad \alpha_1, \alpha_2 < 1 \,.
	\end{cases}
\end{align}
Note that the expressions for the exponents $\mu_1$ and $\mu_2$ in the second lines in eqs. \eqref{I1} and \eqref{I2} are formally valid regardless whether the $\alpha_1 + \alpha_2$ is less than, equal or greater than $1$,  which implies  that the additional algebraic factor can be an increasing function of $|x|$, be independent of $|x|$ in the borderline case $\alpha_1 + \alpha_2 = 1$, or be a decreasing function of $|x|$, respectively.  
Combining eqs. \eqref{I1} and \eqref{I2}, we arrive at our expression \eqref{k2}.

\begin{figure}[t]
	\centering\includegraphics[width=0.5\textwidth]{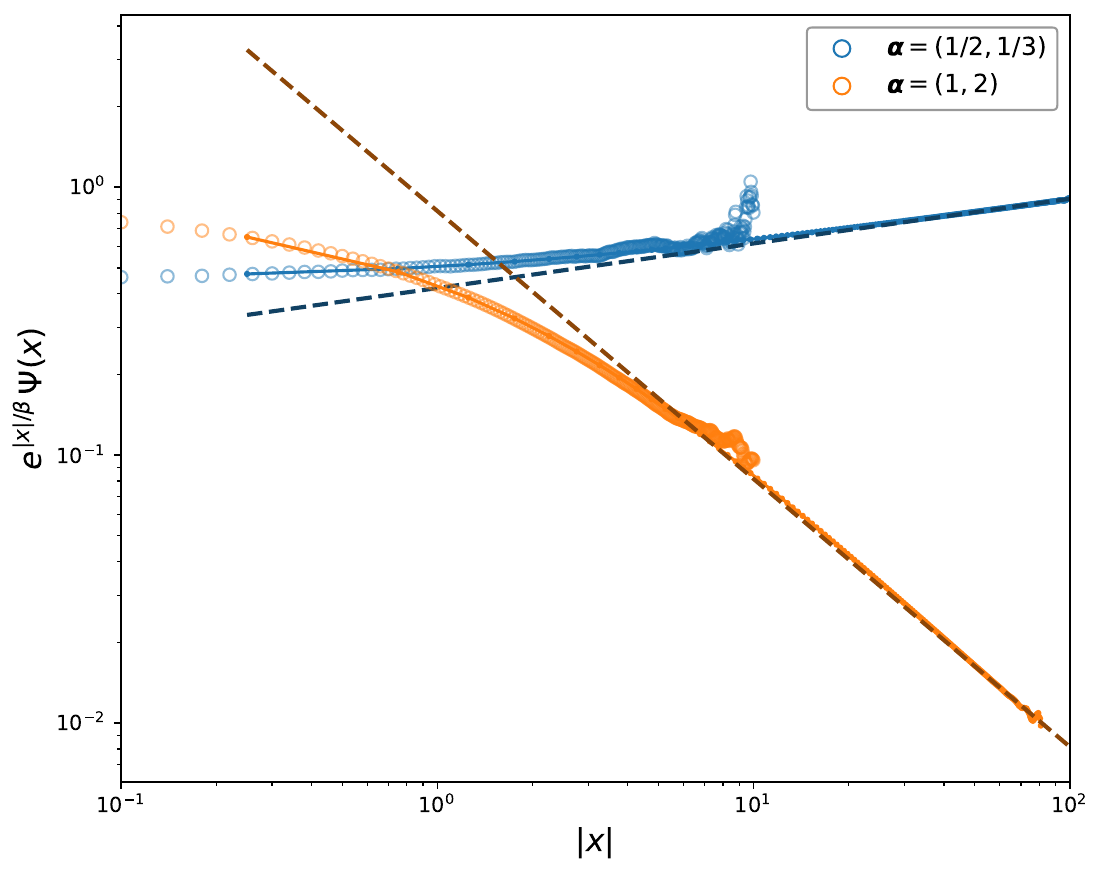}
	\caption{Power-law corrections to the disorder-averaged PDF for $K=2$. 
    The blue and orange open circles correspond to the two cases of random-flight simulations shown in Fig.~\ref{fig:6}, with $\boldsymbol{\alpha}=(1/2, 1/3)$ and $(1, 2)$, respectively. 
    The solid lines represent the numerical evaluation of eq.~\eqref{Psi}, and the dashed lines indicate the predicted power-law asymptotics.}
	\label{fig:S1}
\end{figure}

Next, we explain why we have excluded the cases when either $\alpha_k$ is equal to $1$, while the second is less or equal to $1$. Our exact asymptotic analysis of the integrals in eqs. \eqref{I1} and \eqref{I2} gives for both situations with $\alpha_1 = 1$ and $\alpha_2 < 1$, and $\alpha_1 = \alpha_2 = 1$, the following large-$|x|$ asymptotic behavior
\begin{align}
	\label{log}
	\Psi(x) \simeq \frac{\ln(|x|/\beta)}{|x|^{\alpha_2}} \exp\left( - |x|/\beta\right)   \,,
\end{align}
i.e., 
the leading asymptotics acquires an additional logarithmic correction. The situation will become even more complicated for general $K$ when several $\alpha_k$ will be equal to unity and powers of logarithms will emerge. While these corrections can be analyzed separately, they introduce a number of additional subcases whose classification depends on the multiplicity and arrangement of the marginal exponents. Since such situations form a nongeneric subset of the parameter space and do not modify the main picture, we have excluded them from our analysis. 

The predicted power-law corrections are verified in Fig.~\ref{fig:S1} by comparison with both random-flight simulations and a numerical evaluation of eq.~\eqref{Psi}.
We consider the two parameter sets used in Fig.~\ref{fig:6}, $\boldsymbol{\alpha}=(1/2, 1/3)$ and $(1, 2)$, for which the predicted algebraic factors are $|x|^{1/6}$ and $|x|^{-1}$, respectively.
In both cases, the simulation data and numerical results converge to the predicted power-law behavior at large $|x|$.
The deviations of the simulation data in the far tail are attributed to limited sampling statistics.

We now derive the algebraic prefactors in the leading exponential large-$|x|$ asymptotics in the general case of a superposition of $K$ independent dichotomous noises.   We restrict our attention to the following two classes of reduced switching rates:
(i) $m$ (with $1 \leq m \leq K$) switching  rates satisfy $\alpha_k < 1$, while 
the remaining $K - m$ rates $\alpha_k$ are greater than $1$; (ii) either $\alpha_k > 1$ for all $k=1, 2, \ldots, K$, or there exists a unique index $j$ such that $\alpha_j=1$ and $\alpha_k > 1$ for all $k \neq j$.

We focus on eq.~\eqref{Psi} and rewrite the kernel $Z(\nu)$ in eq. \eqref{Znu} in the form
\begin{align}
	\label{prod}
	Z(\nu) = \prod_{k=1}^K \frac{\,_2F_1\left(1/2, \alpha_k - 1/2; \alpha_k + 1/2; \frac{\delta - 1}{\delta}\right)}{\delta^{1/2}} \,, \qquad \delta =  1 + \beta^2 \nu^2 \,.
\end{align}
Consider next which terms of the kernel exhibit a singular behavior in the complex $\nu$-plane.
Notice that each factor in eq.~\eqref{prod} possesses singular points at $\nu = \pm i/\beta$, at which the denominator vanishes, i.e., $\delta$ becomes equal to zero.  
In the limit $\delta \to 0$, the argument of the Gauss hypergeometric function diverges, such that behavior of each term in the product obtains from the large argument continuation formula for Gauss hypergeometric function:
\begin{align}
	\begin{split}
		\label{exp}
		&\frac{\,_2F_1\left(1/2, \alpha_k - 1/2; \alpha_k + 1/2; \frac{\delta - 1}{\delta}\right)}{\delta^{1/2}} =  D_k  + E_k \, \delta^{\alpha_k - 1} + \text{vanishing when $\delta \to 0$ terms} \,,\\
		&D_k = \frac{\Gamma(\alpha_k + 1/2)}{(\alpha_k - 1) \Gamma(\alpha_k - 1/2)}\,, \quad E_k = \frac{\Gamma(\alpha_k + 1/2) \Gamma(1 - \alpha_k)}{\sqrt{\pi}} \,.
	\end{split}
\end{align}
Note first that this formula is valid excluding certain exceptional values of $\alpha_k$.
The first exceptional value is $\alpha_k= 1/2$. In this case the second
hypergeometric parameter vanishes and one has the exact identity
${}_2F_1\!\left(\frac12,0;1;u\right)\equiv 1$. 
The generic continuation formula therefore degenerates
to give
\begin{align}
	\frac{\,_2F_1\left(1/2, 0; 1; \frac{\delta - 1}{\delta}\right)}{\delta^{1/2}} \equiv   \frac{1}{\delta^{1/2}} \,, \quad \alpha_k = 1/2 \,.
\end{align}
However, that this
degeneracy is completely harmless for our purposes because the resulting
factor is simply
$\delta^{-1/2}$,
which is precisely the singular behaviour predicted by eq. \eqref{exp} for the exponent
$\alpha_k-1=-1/2$ (note that $E_k = 1$ in this case). Thus the coefficient structure changes, but the
branch-point exponent governing the large-$|x|$ asymptotics remains exactly the same. Therefore, the dominant singular behavior when $\delta \to 0$ 
of each factor with $\alpha_k < 1$ (including $\alpha_k = 1/2$) is given by
\begin{align}
	\frac{\,_2F_1\left(1/2, \alpha_k - 1/2; \alpha_k + 1/2; \frac{\delta - 1}{\delta}\right)}{\delta^{1/2}} \simeq E_k \, \delta^{\alpha_k - 1} \,. 
	\end{align}
The second exceptional value is $\alpha_k=1$. The standard connection formula breaks down and
must be replaced by a limiting form involving a logarithm in the leading order
\begin{align}
	\label{log}
		\frac{\,_2F_1\left(1/2, 1/2; 3/2; \frac{\delta - 1}{\delta}\right)}{\delta^{1/2}} = - \frac{1}{2} \ln(\delta/4) + O\left(\delta \ln(1/\delta)\right) \,, \quad \alpha_k = 1 \,.
	\end{align}
This is the most dangerous case in which the hypergeometric function diverges logarithmically when $\delta \to 0$ and hence, may potentially lead to 
logarithmic corrections to the large-$|x|$ asymptotics, as we have observed in the $K = 2$ case.

Finally, when $\alpha_k$ is an integer greater than one, the exponent
difference $1-\alpha_k$ is again an integer. In this case the coefficient $E_k$ diverges, (while $D_k$  remains finite), such  that the connection
formula \eqref{exp} must formally be replaced by its limiting version. For instance, for $\alpha_k = 2$ and $\alpha_k = 3$ we have 
\begin{align}
	\begin{split}
		\label{nlog}
		&\frac{\,_2F_1\left(1/2, 3/2; 5/2; \frac{\delta - 1}{\delta}\right)}{\delta^{1/2}} = \frac{3}{2} + \frac{3}{4} \ln(\delta) \,  \delta + O\left(\delta\right) \,, \qquad \qquad \,\, \alpha_k = 2 \,,\\
		&\frac{\,_2F_1\left(1/2, 5/2; 7/2; \frac{\delta - 1}{\delta}\right)}{\delta^{1/2}} = \frac{5}{4} - \frac{5}{8} \delta  - \frac{15}{16}  \ln(\delta) \,  \delta^2 + O\left(\delta^2\right)\,, \quad \alpha_k = 3 \,.
		\end{split}
	\end{align}
Therefore,  positive integer $\alpha_k >1$ do not require a
separate asymptotic classification, even though the connection
formula \eqref{exp} is no longer applicable. In this case, the constant term $D_k$ is well-defined and the 
first non-analytic term has the form $\ln(\delta) \delta^{\alpha_k - 1}$ for any integer $\alpha_k = n$.
This is the only term which has an impact on the large-$|x|$ asymptotic behavior.

Case (i). Assume that $m \geq 1$ out of $K$ reduced switching rates $\alpha_l < 1$, while the remaining $K - m$ exceed unity. Taking into account that for each factor with $\alpha_k > 1$,  
\begin{align}
	&\frac{\,_2F_1\left(1/2, \alpha_k - 1/2; \alpha_k + 1/2; \frac{\delta - 1}{\delta}\right)}{\delta^{1/2}} = D_k + o(1) \,, 
\end{align}
the kernel in the Fourier integral in eq. \eqref{Psi} attains, in the leading in the limit $\delta \to 0$ order, the following form
\begin{align}
	\begin{split}
	\label{kernel}
	Z(\nu) &= \left(\prod_{\alpha_k > 1} D_k \prod_{\alpha_k < 1} E_k\right) \delta^{\sum_{\alpha_k < 1} \alpha_k - m}  \\&+ \text{less singular  when $\delta \to 0$ terms}\,,
	\end{split}
\end{align}
where the products with the subscripts $\alpha_k < 1$ and $\alpha_k > 1$ extend over all reduced switching rates which are less or are greater than $1$, respectively. There are two different situations which are to be considered separately:

Weakly singular limit. Consider the situation  
\begin{align}
	\sum_{\alpha_k < 1} \alpha_k - m > -1 \,,
	\end{align}
which case is realized when all $\alpha_k$ are only slightly less than $1$ such that the singularity is effectively weak and integrable. Indeed, in this limit the kernel in eq. \eqref{kernel} diverges as a power law $1/\delta^{z}$ with $0 < z < 1$ as $\delta \to 0$.
Applying next the  standard Watson-Darboux theorem (see, e.g., \cite{bleistein}), we find that in this limit
\begin{align}
	\label{dom1}
	\Psi(x) \simeq \frac{1}{|x|^{\sum_{\alpha_k < 1} \alpha_k - m+1}} \exp\left(- |x|/\beta\right) \,.
	\end{align}
Here, the exponent $\mu = \sum_{\alpha_k < 1} \alpha_k - m + 1$ (see eq. \eqref{K}) is greater than zero, and hence, the algebraic prefactor accelerates the exponential decay of the disorder-averaged PDF. 

Strongly singular limit. This case is realized when all $\alpha_k$ ($\alpha_k < 1$) are sufficiently close to zero. In this situation 
\begin{align}
	\sum_{\alpha_k < 1} \alpha_k - m \leq -1 \,,
\end{align}
and may attain large negative values, depending on the actual value of $m$ (and $K$). 
In this case the kernel is strongly divergent, $Z(\nu) \simeq 1/\nu^{z}$ with $z \geq 1$. This means that the standard Watson-Darboux theorem is not applicable in this situation and we have to resort to a different approach which is, in fact, quite straightforward. Recalling the definition of $\delta$ in eq. \eqref{prod} and the definition of $\Psi(x)$, we use directly the result for the kernel in eq.~\eqref{kernel} to find that $\Psi(x)$ obeys
\begin{align}
	\begin{split}
	\Psi(x) &= \frac{\left(\prod_{\alpha_k > 1} D_k \prod_{\alpha_k < 1} E_k\right)}{\pi} \int^{\infty}_0 \frac{\cos(\nu x) d\nu}{\left(1 + \beta^2 \nu^2\right)^{m - \sum_{\alpha_k < 1} \alpha_k}} \\
	&+ \text{correction terms} \,.
	\end{split}
	\end{align}
Performing the integral in the above expression, 
we get
\begin{align}
	\begin{split}
\Psi(x) &= \frac{\left(\prod_{\alpha_k > 1} D_k \prod_{\alpha_k < 1} E_k\right) }{2 \sqrt{\pi} \, \Gamma(m - \sum_{\alpha_k < 1} \alpha_k) \, \beta} 	\left(\frac{|x|}{2 \beta}\right)^{m - \sum_{\alpha_k < 1} \alpha_k} K_{1/2 - m + \sum_{\alpha_k < 1} \alpha_k}\left(\frac{|x|}{\beta}\right) \\
&+ \text{correction terms} \,, 
\end{split}
	\end{align}
which yields the following large-$|x|$ asymptotic form
\begin{align}
	\label{dom2}
	\Psi(x) \simeq |x|^{m - \sum_{\alpha_k < 1} \alpha_k - 1} \exp\left(- |x|/\beta\right) + \text{correction terms} \,.
	\end{align}
Inspecting next the contribution of the correction terms which stem from the representation of the kernel in eq. \eqref{kernel}, one readily finds that they all decay exponentially with $|x|$ and this exponential function is multiplied by a power-law function with the exponent which is smaller than $m - \sum_{\alpha_k < 1} \alpha_k - 1$, which implies that the correction terms provide only a subdominant contribution to the leading term in eq. \eqref{dom2}. Combining eqs. \eqref{dom1} and \eqref{dom2}, we arrive at our result in the first line in eq. \eqref{K}.

Case (ii). Consider next the case when either all $\alpha_k > 1$,  or just one of them is equal to $1$ while the other are greater than $1$. 
Retaining only the leading non-analytic terms in the expansion \eqref{exp}, we obtain
\begin{align}
	\label{last}
	Z(\nu) = \left(\prod_{k = 1}^K D_k\right) +  \sum_{k=1}^K \left(\prod_{\substack{k'=1\\k'\neq k}}^K D_{k'}\right) E_k \delta^{\alpha_k - 1} + \text{higher-order terms}.
\end{align}	
The omitted terms involve higher powers of $\delta$ and are asymptotically subdominant. 

Equation \eqref{last} is directly applicable when none of the $\alpha_k$ are integers. In this case the leading singularities are of the form $\delta^{\alpha_k-1}$ and the Watson--Darboux theorem yields
\begin{align}
	\delta^{\alpha_k-1}
	\, \longrightarrow \, 
	\Psi(x)
	\simeq
	\frac{e^{-|x|/\beta}}{|x|^{\alpha_k}}.
\end{align}
Consequently, the dominant contribution is determined by the smallest value of $\alpha_k$, which immediately leads to the result stated in the second line of eq. \eqref{K}.
\begin{figure}[t]
	\centering\includegraphics[width=\textwidth]{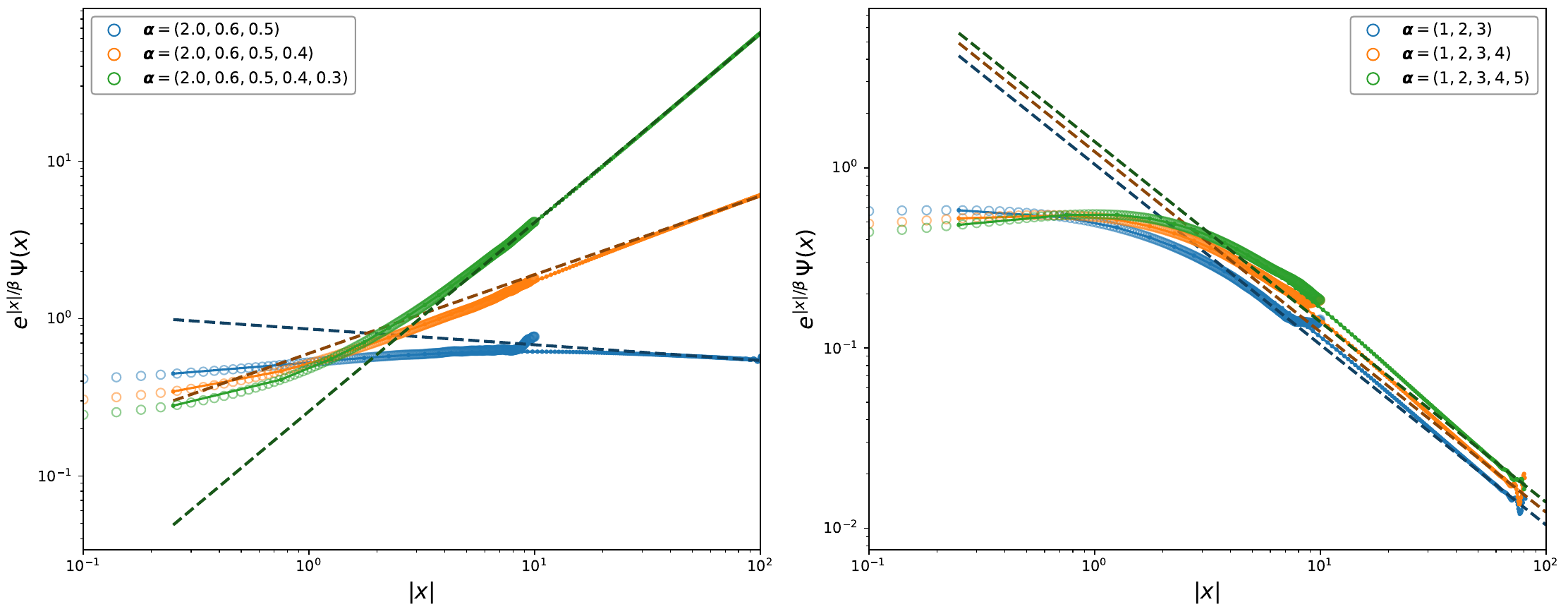}
	\caption{Power-law corrections to the disorder-averaged PDF for $K=3, 4, 5$.
    In the left panel, the blue, orange, and green open circles correspond to the three cases of random-flight simulation shown in the first row of Fig.~\ref{fig:7} (case (i)), with $\boldsymbol{\alpha}=(2.0, 0.6, 0.5)$, $(2.0, 0.6, 0.5, 0.4)$, and $(2.0, 0.6, 0.5, 0.4, 0.3)$, respectively.
    The corresponding open circles in the right panel represent the three cases shown in the second row of Fig.~\ref{fig:7} (case (ii)), with $\boldsymbol{\alpha}=(1.0, 2.0, 3.0)$, $(1.0, 2.0, 3.0, 4.0)$, and $(1.0, 2.0, 3.0, 4.0, 5.0)$, respectively.
    The solid lines represent the numerical evaluation of eq.~\eqref{Psi}, and the dashed lines indicate the predicted power-law asymptotics.}
	\label{fig:S2}
\end{figure}

Suppose next that some of the exponents $\alpha_k$ are integers greater than $1$. In this case the corresponding non-analytic terms are not $\delta^{\alpha_k-1}$ but rather $\delta^{\alpha_k-1}\ln\delta$ (see eq.~\eqref{nlog}). The discontinuities of the terms at the cut are equal to	$2\pi i\,\delta^{n-1}$, and hence, 
their contribution to the large-$|x|$ asymptotics is
\begin{align}
	\label{100}
	\delta^{n-1}\ln\delta
	\quad\longrightarrow\quad
	\Psi(x)
	\simeq
	\frac{e^{-|x|/\beta}}{|x|^{n}}.
\end{align}
Therefore integer values $\alpha_k=n>1$ produce exactly the same asymptotic power as obtained by analytic continuation of the non-integer result.

Finally, suppose that exactly one exponent is equal to unity. The corresponding factor behaves as $\ln\delta$ (see eq.~\eqref{log}), and
\begin{align}
	\ln\delta
	\quad\longrightarrow\quad
	\Psi(x)
	\simeq
	\frac{e^{-|x|/\beta}}{|x|}.
\end{align}
Since all remaining factors are finite as $\delta\to0$, this contribution dominates the large-$|x|$ behavior. If some of the remaining exponents are integers greater than $1$, products involving higher powers of $\ln\delta$ may appear. Such terms contribute only to subleading orders and do not affect the leading asymptotic behavior.

Collecting the above results, we recover the asymptotic form stated in the second line of eq.~\eqref{K}.

The predicted power-law corrections for $K=3,4,$ and $5$ are verified in Fig.~\ref{fig:S2} by comparison with both random-flight simulations and numerical evaluations of eq.~\eqref{Psi}.
For case (i), corresponding to the first row of Fig.~\ref{fig:7}, the predicted algebraic factors are $|x|^{-0.1}$, $|x|^{0.5}$, and $|x|^{1.2}$ for $K=3$, $4$, and $5$, respectively.
For case (ii), corresponding to the second row of Fig.~\ref{fig:7}, the predicted algebraic factor is $|x|^{-1}$ for all three values of $K$.
In all cases, the simulation data and numerical results approach the predicted power-law asymptotics at large $|x|$, confirming the analytical predictions.
The deviations of the simulation data at the largest $|x|$ are expected to arise from limited sampling statistics in the far tail.

\end{document}